\documentclass[11pt,a4paper,twoside]{article}
\usepackage[all,2cell,v2]{xy}
\usepackage{axodraw}
\usepackage{a4wide}
\usepackage{fancyvrb}
\usepackage{subfigure}
\usepackage{url}
\usepackage{calc}
\usepackage{latexsym}
\usepackage{amsmath}
\usepackage{amsfonts}

\newtheorem{theorem}{Theorem}

\newtheorem{prop}[theorem]{Proposition}

\newenvironment*{codeblock}%
  {\Verbatim[fontsize=\small,%
             numbers=left,%
             baselinestretch=0.9,%
             xleftmargin=2em,%
             numbersep=1em,
             commandchars=\\\{\}]}
  {\endVerbatim}

\newcommand{\eps}{\varepsilon}         
\newcommand{\Id}
           {1\hspace{-0.5mm}{\begin{picture}(10,10)(0,0)
                               \SetWidth{0.8}\Line(0,0.2)(0,8)
                               \SetWidth{0.3}\Line(-3.8,0.14)(1.8,0.14)
                             \end{picture}}\hspace{-2.25mm}}

\author{Isabella Bierenbaum\footnote{e-mail: {\tt bierenbaum@thep.physik.uni-mainz.de}},
Richard Kreckel\footnote{e-mail: {\tt
richard.kreckel@uni-mainz.de}}\\{\tiny Department of Physics,
Mainz University, D-55099 Mainz, Germany}\and  Dirk
Kreimer\footnote{e-mail: {\tt dkreimer@bu.edu}}\\
{\tiny Mathematics Department, Boston University, Boston MA02215,
USA}}

\date{November 21, 2001}

\title{On the Invariance of Residues of Feynman Graphs}

\begin{document}
\maketitle \vspace{-24em}\begin{flushright}
MZ-TH/01-33\\
BUCMP/01-01\\
arXiv.org: hep-th/0111192
\end{flushright}\vspace{16em}

\begin{abstract}
We use simple iterated one-loop graphs in massless Yukawa theory
and QED to pose the following question: what are the symmetries of
the residues of a graph under a permutation of places to insert
subdivergences. The investigation confirms partial invariance of
the residue under such permutations: the highest weight
transcendental is invariant under such a permutation. For QED this
result is gauge invariant, i.e.\ the permutation invariance holds
for any gauge. Computations are done making use of the Hopf
algebra structure of graphs and employing GiNaC to automate the
calculations.
\end{abstract}

\pagenumbering{arabic}

\section{Introduction}
This paper serves three purposes: i) it employs
GiNaC~\cite{GiNaCpaper} in Feynman diagram calculations and provides
algorithms which automate the renormalization process, very much in
the spirit of~\cite{BrKr}; ii) it investigates symmetries of short
distance singularities under permutations of places where to insert
subdivergences in a graph; iii) it once more confirms the presence or
absence of transcendental coefficients of short-distance singularity
in accordance with the topology of a graph.

Our laboratory of investigation are simple one-loop Feynman graphs
in massless Yukawa theory or QED, inserted into each other in
tree-like hierarchies. Thus, the combinatorics of renormalization
boils down to the Hopf algebra of decorated rooted trees with only
a small number of decorations and the analytical challenge posed
by those decorations reduces to expansions of $\Gamma$-functions
near unit argument. The question we can ask is for the
distribution of the Riemann $\zeta$-function over the various
poles in the Laurent series of graphs of that form.

In contrast to~\cite{BrKr} and subsequent papers~\cite{BrKrff},
where the renormalization problem was automated in a similar
context optimized for speed and efficiency, we have developed here
algorithms which allow for non-trivial spin-structures and an easy
generalization to arbitrary decorations: the primitive decorations
can be inserted as arguments so that the algorithm can handle
arbitrary primitive graphs when their analytic structure becomes
known.

We work in the context of dimensional regularization, so that any
Feynman graph becomes a Laurent series in $\eps=(4-D)/2$, the
deviation of the dimension from its integer value four, and the
pole terms reflect the short distance singularities in the theory.
The first order pole is denoted as the residue of a graph. Its
significance lies in the fact that higher pole terms can be
reduced to polynomials in residues~\cite{RHII}, and in the fact
that the residue of a primitive graph is an invariant under
diffeomorphisms of external parameters of the graph
(diffeomorphisms of external momenta and masses) as well as an
invariant under variations of renormalization schemes. Such a
residue is a motivic number then in some modern mathematic
parlance. The question as to which class of such numbers is
sufficient to describe the residues of a given quantum field
theory is open and fascinating~\cite{review}.

We only study two much more basic questions, motivated by previous
and ongoing investigations into the analytic structure of pole terms
and residues in particular.

The first is the independence of the appearance of transcendentals
under variations of the quantum field theory which realizes a
graph with a given topology. To specify the topology of a
one-particle irreducible Feynman graph $\Gamma$, let us consider
the adjacency matrix $M(\Gamma)$ of $\Gamma$. If $\Gamma$ has $n$
vertices, this matrix is a $n\times n$ matrix. We take for its
non-zero entries pairs $\textsf{(propagator type, powercounting
weight)}$, ie.\ each non-zero entry $M(\Gamma)_{ij}$ specifies
that vertex $i$ is connected to vertex $j$ by a propagator of some
type, which has a certain powercounting weight.\footnote{We can
notate the type of vertex in the diagonal entries of this
symmetric matrix (no vertex is connected to itself by a
propagator).} In the cases studied here, the possible entries are
$$ \textsf{(fermion,1),(photon,2),(scalar boson,2)}.$$ The
listing of the powercounting weight is redundant, as it is
determined by the type of the propagator. We list it just for easy
reference.

The graphs of QED and Yukawa theory which we will compare always
have adjacency matrices which agree in all their zero entries,
$$ M(\Gamma_{\rm QED})_{ij}=0\Leftrightarrow M(\Gamma_{\rm Yuk})_{ij}=0 $$
 and
agree for each non-zero entry in the powercounting degree of the
corresponding edge. It is only the nature of the edges which changes
from the spin one vector boson propagator --the QED photon-- to the
spin zero scalar boson propagator in Yukawa theory. Note that the
structure of short distance singularity then remains fixed in the
transition of one theory to the other.  We then expect and confirm
that rational numbers can vary in the transition from one theory
to the other, while the transcendentals we see remain invariant
and specific for the chosen topology. Here, a topology is uniquely
described by considering in such an adjacency matrix all non-zero
entries as equal. So it just gives information about how vertices
are connected, but forgets about the nature of the propagators
establishing that connection. To find non-rational numbers, we
have to go up to four loops at least in the simple class of
Feynman graphs which we consider. There, it is the swiss cheese
topology of Fig.1a in which we expect to see a residue
$\sim\zeta(3)$, while the ladder topology of Fig.1b are known to
have only rational residues~\cite{Rat}.

The second question is of different nature: for a log-divergent vertex
graph which is primitive under the coproduct, it is evident that its
residue has the above mentioned invariances. The Hopf algebra
structure immediately allows to prove that in a vertex graph which is
not primitive under the coproduct such that the graph contains
divergent subgraphs, the coefficient of the highest order pole still
has these symmetries and is given by an easy calculable product of
residues with a combinatorial factor, determined by the scattering
type formula of~\cite{RHII}, which incorporates the 't~Hooft relations
between higher pole terms in graphs.

What we call here the 't~Hooft relations in accordance
with~\cite{RHII} is the simple fact that higher order poles in a graph
$\Gamma$ with UV-divergent subgraphs $\gamma$ can be calculated from
products of lower order poles of these subgraphs $\gamma$ and their
complements $\Gamma/\gamma$. This is well-known and a necessary
requirement to make the renormalization group work: the $Z$-factor for
a given physical quantity is an invertible (it starts with 1) formal
series over counterterms of graphs such that its logarithmic
derivative with respect to the variation of the renormalization scale
$\nu$, $d\log(Z)/d\nu^2$, is finite.  This finiteness establishes
relations between pole terms which, for the case of dimensional
regularization, were, it seems, first explored by 't~Hooft. These
relations are a direct consequence of the mathematical structure of
the Hopf algebra underlying renormalization, and its one-parameter
group of automorphisms~\cite{RHII}.

Let us now consider the residue of such a graph which does have
divergent subgraphs. Typically, this residue will be a number
which can be decomposed in terms of transcendental weight: it will
contain contributions ranging from rational numbers to monomials
in $\zeta(j)$ of up to transcendental weight $l-1$, where $l$ is
the bidegree of the graph, calculated from its
coproduct~\cite{Bigrad}\footnote{For all graphs considered here, it simply
agrees with the number of loops in Yukawa theory. For the QED
vertex, a shift by one unit can appear as explained later in the
text.}. We cannot expect the whole residue to be invariant under
the above symmetries, as rational numbers can and will vary
freely. But here we report a remarkable partial symmetry observed
in our rather restricted lab of iterated one-loop graphs: the
highest weight transcendental in the residue is invariant under
permutations of external momentum as described below. We confirm
this by empirical calculation to high loop orders. We finally
prove the result in the context of the simple iterated one-loop
graphs considered here. One has an almost elementary proof in this
context and we will discuss the difficulties which arise in the
general case.
 The nature of
this result fits nicely with a structural investigation of
Dyson-Schwinger equations to be delivered
elsewhere~\cite{DirkNew}. Note again that changing the momentum flow but
maintaining the topology corresponds to alterations of the type of
non-zero entries in a suitable adjacency matrix, as
Fig.\ref{fig:two_vertex_types} clearly exhibits. Again, the degree
of powercounting and hence the structure of short distance
singularities, as well as the topology, remain unchanged.

To understand the type of symmetry we want to investigate, let us
consider the one-loop vertex function. If $\Gamma(p_1,p_2,p_3)$ is
a vertex correction with momentum $p_1$ for the external boson,
$p_2$ for the incoming fermion and $p_3=p_1+p_2$ for the outgoing
fermion, then we will compare $\Gamma^1:=\Gamma(0,p,p)$
with $\Gamma^2:=\Gamma(p,p,0)$
at the one-loop level. It is a single permutation of the flow of
an  external momentum  for a vertex function at zero momentum
transfer, which distinguishes the two one-loop functions. Locality
of counterterms ensures that the counterterm for a vertex
correction graph is invariant under this permutation for any
graph. So nothing exciting can be learned from just changing the
momentum flow in a given vertex graph. Now let us start with such
one-loop vertex corrections and let us insert more one-loop vertex
corrections always at the vertex of zero momentum transfer (zmt),
which indeed changes under the above permutation. This gives the
topological equivalent Feynman graphs $\Gamma^1,\Gamma^2$, with
permuted types of propagators, of Fig.\ref{fig:two_vertex_types}.
They indeed have adjacency matrices such that again
$$M(\Gamma^1)_{ij}=0\Leftrightarrow M(\Gamma^2)_{ij}=0.$$

Note that it is the requirement to maintain the same topology
which forces us to consider quite different looking graphs. For
them, the usual field theoretic requirements only demand that the
coefficients of the highest order poles are identical. All
non-leading pole terms in $\eps$ do not have to and indeed do not
agree.

But a closer investigation will establish a remaining partial
symmetry between those graphs, such that the highest weight
transcendental in the residue is invariant. Such symmetries are of
great interest. Feynman graphs with subdivergences can be built by
using the pre-Lie structure of graphs, which results from
underlying insertion operads~\cite{review}. The question to what
extent the permutation group acts trivially under such insertions
is a natural question in operad theory which needs to be answered
to understand these operads. It  directly leads to the questions
studied here in a simplified context. We regard the results
reported here as a first step in a detailed analysis of actions of
the permutation group in this context.

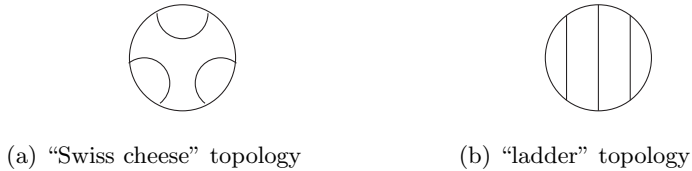
\begin{figure}[t]
\begin{center}
  \subfigure[``Swiss cheese'' topology]{ \SetScale{0.8}
  \begin{picture}(150,30)(-50,0)
    \BCirc(40,20){25}
    \CArc(40,41)(12,180,0)
    \CArc(58,8)(12,52,-128)
    \CArc(22,8)(12,-52,128)
  \end{picture}}
  \subfigure[``ladder'' topology]{
  \begin{picture}(150,30)(-50,0)
    \SetScale{0.8}
    \BCirc(40,20){25}
    \Line(40,-5)(40,45)
    \Line(55,0)(55,40)
    \Line(25,0)(25,40)
  \end{picture}}
\end{center}\vspace{-2em}
\caption{\small ``Swiss cheese'' and ``ladder'' master topology.}
\label{fig:ladder_and_cheese}
\end{figure}

With regard to the first question, we remind the reader that for a
massless Yukawa theory represented as a Laurent-series in $\eps$,
it was already shown in~\cite{On_knots} that graphs with the swiss
cheese topology of figure~\ref{fig:ladder_and_cheese}.a possess a
$\zeta(n)$ in their counterterm, whereas graphs with a ladder
topology like in figure~\ref{fig:ladder_and_cheese}.b just
evaluate to rational coefficients. Similar results were obtained
in QED and other theories for vertex functions of type
$\Gamma(0,p,p)$ and self-energies, in~\cite{BrKr,BrKrff,Rat}.

Our first aim is now to find out if this remains true in massless
QED for both types of momentum flow and if there are any
symmetries in the coefficients of the $\zeta$-functions in both
theories, under the permutation between the two types of vertex
functions $\Gamma^1$ and $\Gamma^2$. Therefore we will rebuild the
scheme given in~\cite{On_knots} for the massless Yukawa theory
with an extension to vertex corrections that carry a different
flow of momentum in the sense described above, and similarly for
QED using the matrix calculus of~\cite{Delbourgo}. The resulting
algorithm is implemented using the GiNaC library and will be
described in section~\ref{sec:implementation}. With the help of
this program, we will calculate the antipode of graphs that
represent the different topologies given above and compare them.

\section{Calculations}
Consider a one-loop contribution to the fermionic propagator:
\begin{center}
\begin{tabular}{lcl}
\SetScale{0.7}
\begin{picture}(130,50)(0,0)
\ArrowLine(40,-10)(120,-10)
\ArrowLine(0,-10)(40,-10)\ArrowLine(120,-10)(160,-10)
\DashArrowArc(80,-10)(40,0,180){4}
\LongArrowArc(80,-10)(25,60,120)
\Vertex(120,-10){1.5}\Vertex(40,-10){1.5}
\PText(20,1)(0)[]{q}
\PText(140,1)(0)[]{q}
\PText(80,50)(0)[]{k}
\PText(80,-20)(0)[]{q+k}
\end{picture}
\SetScale{1} & $ \equiv $ & $ \Sigma_{[0,0]}(q^2) \label{Sigm0} =
\displaystyle{\int} d^Dk \,
\frac{1}{(q\hspace{-2mm}{/}+k\hspace{-2mm}{/})} \frac{1}{k^2}
$\\
\vspace{0.2cm}
\end{tabular}
\end{center}
where the subscript ``$[0,0]$'' will soon serve to give the number
of one-loop subdivergences to be inserted inside the diagram at
the fermionic or bosonic line. All integrals considered in this
paper can be reduced to integrals $F_{a,b}$~\cite{Pascual}:\footnote{These
$\Gamma$-functions will give rise
  to the $\zeta$-functions we are looking for.  An easy way to see
  the connection between the $\Gamma$- and the $\zeta$-function is the
  formula:
\begin{eqnarray*}
\Gamma\,(1+\eps) \label{Gammaformel} &=& \exp\,(-\gamma\eps)\:
\exp\,\left(\sum_{n=2}^{\infty}\frac{\zeta(n)}{n}(-\eps)^n\right)\,
,\quad\quad |\eps|<1\end{eqnarray*} Internally, GiNaC follows a
different approach.  It computes the derivatives
$\Gamma'(x)=\Gamma(x)\psi(x)$ and
$\frac{d^n}{dx^n}\psi(x)=\psi_n(x)$ in terms of polygamma
functions $\psi_m(x)$ and ``knows'' how to evaluate polygamma
functions at integer arguments, e.g.\ $\psi(1)=-\gamma$ and
$\psi_n(1)=(-)^{n+1}n!\zeta(n+1)$.}
\begin{eqnarray}
I(q;a,b) \label{PasTar} &\equiv& \frac{1}{(\nu^2)^{-\eps}} \int
d^Dk
\frac{1}{[k^2]^{a}[(q-k)^2]^{b}}\nonumber\\
&=:& \frac{1}{(\nu^2)^{-\eps}} [q^2]^{(2-(a+b)-\eps)} F_{a,b}
\end{eqnarray}
with
\begin{eqnarray}
\label{FFkt2} F_{a,b}\equiv F_{b,a} &:=& \frac
{\Gamma\,(2-a-\eps)\:\Gamma\,(2-b-\eps)\:\Gamma\,(a+b-2+\eps)}
{\Gamma\,(a)\Gamma\,(b)\Gamma\,(4-a-b-2\eps)},
\end{eqnarray}
which we typically need for $a=n_1+n_2\eps,b=m_1+m_2\eps$, for
integers $n_1,n_2,m_1,m_2$. This reduces the identification of
non-rationals (transcendentals, we dare say in the following) to an
expansion of the $\Gamma$-function near unit argument, as
promised. Note that
 \begin{eqnarray}
\label{Tadpole} F_{a,-n}\,&=&\,F_{-n,b}\,=\,0
\quad,\quad\quad
n\in \mathbb{N}\:^0.
\end{eqnarray}
Accordingly, in our conventions
\begin{eqnarray}
\label{Sigma00}
\Sigma_{[0,0]}(q^2)=[q^2]^{-\eps}
\frac{1}{2}F_{1,1}\;q\hspace{-2mm}{/}=:[q^2]^{-\eps}\Sigma_{0,0}\;q\hspace{-2mm}{/}.
\end{eqnarray}


\subsection{Yukawa theory}
The study of iterated one-loop integrals reduces to the study of the
following elementary functions, which we will call {\em characterizing
functions}: The one-loop fermion self-energy with insertions at its
two internal propagators demands knowledge of
\begin{eqnarray}
\Sigma_{i,j} \label{SY} &:=&
\frac{1}{2}[F_{i\varepsilon,1+j\varepsilon}+F_{1+i\varepsilon,1+j\varepsilon}-F_{1+i\varepsilon,j\varepsilon}].
\end{eqnarray}
The one-loop boson self-energy with insertions at its two internal
edges is given by $\Pi_{i,j}$ and the two one-loop vertex functions
at zero momentum transfer with insertions at either the zero momentum
vertex, or internal lines, need knowledge of functions $\Gamma_{i,j}$
for the vertex corrections.
\begin{eqnarray}
\Pi_{i,j}/[\textrm{tr}(\Id)] \label{PY} &:=&
-\frac{1}{2}[F_{i\varepsilon,1+j\varepsilon}-F_{1+i\varepsilon,1+j\varepsilon}+F_{1+i\varepsilon,j\varepsilon}]
\\
\Gamma^1_{i,j}
&:=&
 F_{1+i\eps,1+j\eps}
\\
\Gamma^2_{i,j} &:=&
\frac{1}{2}[F_{2+i\varepsilon,j\varepsilon}-F_{2+i\varepsilon,1+j\varepsilon}+F_{1+i\varepsilon,1+j\varepsilon}].
\end{eqnarray}
Here, we have divided the boson self-energy by the trace of the
unit matrix $\textrm{tr}(\Id)$ (trace over spinorial indices) for
easier comparison of insertions of subgraphs into bosonic and
fermionic lines.
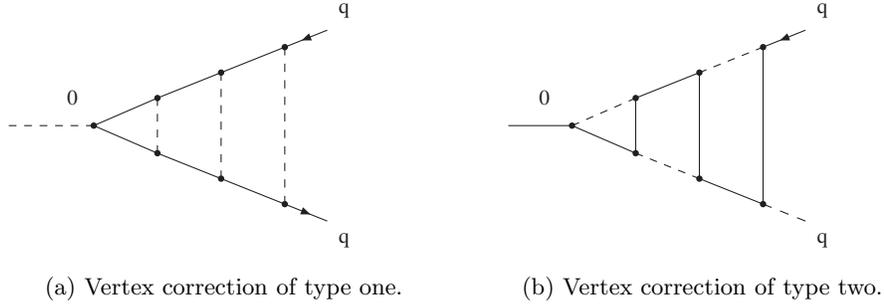
\begin{figure}[t]
\begin{center}
\subfigure[Vertex correction of type one.]{ \SetScale{0.8}
\begin{picture}(170,90)(0,0)
\ArrowLine(150,95)(130,87)
\Line(130,87)(100,75)
\Line(100,75)(70,63)
\Line(70,63)(40,50)
\Line(40,50)(70,37)
\Line(70,37)(100,25)
\Line(100,25)(130,13)
\ArrowLine(130,13)(150,5)
\DashLine(0,50)(40,50){4}
\DashLine(130,87)(130,13){4}
\DashLine(100,75)(100,25){4}
\DashLine(70,63)(70,37){4}
\Vertex(130,87){1.5}\Vertex(130,13){1.5}\Vertex(40,50){1.5}
\Vertex(100,75){1.5}\Vertex(100,25){1.5}
\Vertex(70,63){1.5}\Vertex(70,37){1.5}
\PText(158,108)(0)[]{q}
\PText(158,0)(0)[]{q}
\PText(30,65)(0)[]{0}
\end{picture}}
\subfigure[Vertex correction of type two.]{ \SetScale{0.8}
\begin{picture}(170,90)(0,0)
\ArrowLine(150,95)(130,87)
\DashLine(130,87)(100,75){4}
\Line(100,75)(70,63)
\DashLine(70,63)(40,50){4}
\Line(40,50)(70,37)
\DashLine(70,37)(100,25){4}
\Line(100,25)(130,13)
\DashLine(130,13)(150,5){4}
\Line(10,50)(40,50)
\Line(130,87)(130,13)
\Line(100,75)(100,25)
\Line(70,63)(70,37)
\Vertex(130,87){1.5}\Vertex(100,75){1.5}\Vertex(70,63){1.5}
\Vertex(40,50){1.5}\Vertex(70,37){1.5}
\Vertex(100,25){1.5}\Vertex(130,13){1.5}
\PText(158,108)(0)[]{q}
\PText(158,0)(0)[]{q}
\PText(27,65)(0)[]{0}
\end{picture}}
\end{center}
\vspace{-0.7cm} \caption{\small The two types of vertex
corrections.} \label{fig:two_vertex_types}
\end{figure}

The functions $\Gamma^1$ and $\Gamma^2$ represent the two types of
vertex corrections shown in figure~\ref{fig:two_vertex_types}.

The indices $i$ and $j$ give the number of subdivergences at different lines:

\begin{tabular}{ll}
$\Sigma_{i,j}$:
&
$i$ = number of subdivergences at the fermion line,\\
&
$j$ = number of subdivergences at the boson line.\\
$\Pi_{i,j}$:
&
$i$ = number of subdivergences at the lower fermion line,\\
&
$j$ = number of subdivergences at the upper fermion line.\\
$\Gamma^1_{i,j}$: & $i$ = number of fermion self-energies and
vertex corrections\\ & plugged into the zmt vertex
and the internal edges connected to it,\\
&
$j$ = number of subdivergences at the boson line not connected to this vertex.\\
$\Gamma^2_{i,j}$: & $i$ = number of fermion or boson self-energies
and vertex corrections\\ & plugged into the zmt vertex
and the internal edges connected to it,\\
&
$j$ = number of fermion self-energies at the fermion line not connected to\\
 & the vertex of zmt.\\
\end{tabular}

To denote complete graphs, we introduce additional functions
$\Sigma_{[i,j]}$, $\Pi_{[i,j]}$ and $\Gamma_{[i,j,k]}$ which
notate the different kinds of subdivergences (cf. $\Sigma_{[0,0]}$
in (\ref{Sigma00})). The indices count for $\Sigma_{[i,j]}$ and
$\Pi_{[i,j]}$ as the indices in their corresponding characterizing
functions given above. For vertex corrections of type one and two,
the notation $\Gamma_{[i,j,k]}$ denotes the three types of
insertions in the following way: $i$ insertions of self-energies
at the edges connected to the zero momentum vertex, $j$ insertions
of self-energies at the edge not connected to that vertex, and $k$
insertions of vertex corrections at the zmt vertex. Table
\ref{tab:tabelle_eins} gives examples.

\begin{table}[t]
\begin{tabular*}{\textwidth}{@{}l@{\extracolsep{\fill}}l@{\extracolsep{1ex}}l@{\extracolsep{1ex}}l@{}}\hline
\SetScale{0.3}
\SetWidth{2}
\begin{picture}(55,22)(0,0)
\Line(0,0)(160,0)
\DashArrowArc(80,0)(40,0,180){4}
\Vertex(120,0){1.5}\Vertex(40,0){1.5}
\end{picture}
&
$
\Sigma_{[0,0]}(q^2)
\rule[-4mm]{0mm}{8mm}
$
&
=
&
$
(-ig)^2\,
\Sigma_{0,0}
[q^2]^{-\eps}q\hspace{-2mm}{/}
$
\\[-.5em]
\SetScale{0.3}
\SetWidth{2}
\begin{picture}(55,22)(0,0)
\Line(0,0)(160,0)
\DashArrowArc(80,0)(60,0,180){4}
\DashArrowArc(80,0)(25,0,180){4}
\Vertex(140,0){1.5}\Vertex(20,0){1.5}
\Vertex(105,0){1.5}\Vertex(55,0){1.5}
\end{picture}
&
$
\Sigma_{[1,0]}(q^2)
\rule[-4mm]{0mm}{8mm}$&
=&$
(-ig)^4
\Sigma_{1,0}
\Sigma_{0,0}
[q^2]^{-2\eps}
q\hspace{-2mm}{/}
$
\\[-.5em]
\SetScale{0.18}\SetWidth{2}
\hspace{0.03cm}
\begin{picture}(55,22)(0,0)
\Line(-20,-30)(30,-30)
\Line(30,-30)(60,-30)
\Line(60,-30)(110,-30)
\Line(110,-30)(150,-30)
\Line(150,-30)(200,-30)
\Line(200,-30)(230,-30)
\Line(230,-30)(280,-30)
\DashArrowArc(130,-30)(100,0,180){4}
\DashArrowArc(85,-30)(25,0,180){4}
\DashArrowArc(175,-30)(25,0,180){4}
\Vertex(30,-30){1.5}\Vertex(230,-30){1.5}
\Vertex(60,-30){1.5}\Vertex(110,-30){1.5}
\Vertex(150,-30){1.5}\Vertex(200,-10){1.5}
\BCirc(130,70){22}
\Vertex(108,70){1.5}\Vertex(152,70){1.5}
\end{picture}
&
$
\Sigma_{[2,1]}(q^2)
\rule[-4mm]{0mm}{8mm}
$&=&$
(-ig)^8\,
\Sigma_{2,1}
\Pi_{0,0}
(\Sigma_{0,0})^2
[q^2]^{-4\eps}
q\hspace{-2mm}{/}
$
\\[-.5em]
\SetScale{0.3}\SetWidth{2}
\hspace{0.015cm}
\begin{picture}(55,22)(0,0)
\DashLine(0,20)(45,20){4}
\DashLine(115,20)(160,20){4}
\BCirc(80,20){35}
\end{picture}
&
$
\Pi_{[0,0]}(q^2)
\rule[-4mm]{0mm}{8mm}
$&=&$
(-ig)^2\,
\Pi_{0,0}
[q^2]^{-\eps}q^2
$
\\[-.5em]
\SetScale{0.3}\SetWidth{2}
\hspace{0.02cm}
\begin{picture}(55,22)(0,0)
\DashLine(0,25)(40,25){4}
\DashLine(120,25)(160,25){4}
\BCirc(80,25){40}
\DashArrowArc(79,-3)(28,0,176){4}
\DashArrowArc(80,-13)(14,0,180){4}
\DashArrowArc(80,61)(15,178,360){4}
\end{picture}
&
$
\Pi_{[2,1]}(q^2)
\rule[-4mm]{0mm}{8mm}
$&=&$
(-ig)^8\,
\Pi_{2,1}\Sigma_{1,0}(\Sigma_{0,0})^2[q^2]^{-4\eps}q^2
$
\\[-.5em]
\SetScale{0.3}\SetWidth{2}
\begin{picture}(55,22)(0,0)
\Line(150,65)(125,55)
\Line(125,55)(40,20)
\Line(40,20)(125,-15)
\Line(125,-15)(150,-25)
\DashLine(0,20)(40,20){4}
\DashLine(125,55)(125,-15){4}
\Vertex(125,55){1.5}\Vertex(125,-15){1.5}\Vertex(40,20){1.5}
\end{picture}
&
$
\Gamma^1_{[0,0,0]}(q^2)
\rule[-4mm]{0mm}{8mm}
$
&=&
$
(-ig)^3\,
\Gamma^1_{0,0}[q^2]^{-\eps}
$
\\[-.0em]
\SetScale{0.3}\SetWidth{2}
\begin{picture}(55,22)(0,0)
\Line(150,65)(130,57)
\Line(130,57)(100,45)
\Line(100,45)(70,33)
\Line(70,33)(40,20)
\Line(40,20)(70,7)
\Line(70,7)(100,-5)
\Line(100,-5)(130,-17)
\Line(130,-17)(150,-25)
\DashLine(0,20)(40,20){4}
\DashLine(130,57)(130,-17){4}
\DashLine(100,45)(100,-5){4}
\DashLine(70,33)(70,7){4}
\Vertex(130,57){1.5}\Vertex(130,-17){1.5}\Vertex(40,20){1.5}
\Vertex(100,45){1.5}\Vertex(100,-5){1.5}
\Vertex(70,33){1.5}\Vertex(70,7){1.5}
\end{picture}
&
$
\Gamma^1_{[0,0,2]}(q^2)
\rule[-4mm]{0mm}{8mm}
$
&=&
$
(-ig)^7\,
\Gamma^1_{2,0}\Gamma^1_{1,0}\Gamma^1_{0,0}[q^2]^{-3\eps}
$
\\[.0em]
\SetScale{0.3}
\SetWidth{2}
\begin{picture}(55,22)(4,0)
\Line(150,65)(125,55)
\Line(125,55)(40,20)
\Line(40,20)(125,-15)
\Line(125,-15)(150,-25)
\DashLine(0,20)(40,20){4}
\DashLine(125,55)(125,-15){4}
\Vertex(125,55){1.5}\Vertex(125,-15){1.5}\Vertex(40,20){1.5}
\DashArrowArc(82,3)(22,-22,158){2}
\DashArrowArc(82,3)(10,-22,158){2}
\end{picture}
&
$
\Gamma^1_{[2,0,0]}(q^2)
\rule[-4mm]{0mm}{8mm}
$
&=&
$
(-ig)^7\,
\Gamma^1_{2,0}\Sigma_{1,0}\Sigma_{0,0}[q^2]^{-3\eps}
$
\\[.0em]
\SetScale{0.3}
\SetWidth{2}
\begin{picture}(55,22)(4,0)
\Line(150,65)(125,55)
\Line(125,55)(40,20)
\Line(40,20)(125,-15)
\Line(125,-15)(150,-25)
\DashLine(0,20)(40,20){4}
\DashLine(125,55)(125,-15){4}
\Vertex(125,55){1.5}\Vertex(125,-15){1.5}\Vertex(40,20){1.5}
\BCirc(125,20){14}
\end{picture}
&
$
\Gamma^1_{[0,1,0]}(q^2)
\rule[-4mm]{0mm}{8mm}
$
&=&
$
(-ig)^5\,
\Gamma^1_{0,1}\Pi_{0,0}[q^2]^{-2\eps}
$
\\[.0em]
\SetScale{0.3}\SetWidth{2}
\begin{picture}(55,22)(0,0)
\Line(150,65)(125,55)
\DashLine(125,55)(40,20){4}
\Line(40,20)(125,-15)
\DashLine(125,-15)(150,-25){4}
\Line(0,20)(40,20)
\Line(125,55)(125,-15)
\Vertex(125,55){1.5}\Vertex(125,-15){1.5}\Vertex(40,20){1.5}
\end{picture}
&
$
\Gamma^2_{[0,0,0]}(q^2)
\rule[-4mm]{0mm}{8mm}
$
&=&$
(-ig)^3\,
\Gamma^2_{0,0}[q^2]^{-\eps}
$
\\[.0em]
\SetScale{0.3}
\SetWidth{2}
\begin{picture}(55,22)(4,0)
\Line(150,65)(130,57)
\DashLine(130,57)(100,45){4}
\Line(100,45)(70,33)
\DashLine(70,33)(40,20){4}
\Line(40,20)(70,7)
\DashLine(70,7)(100,-5){4}
\Line(100,-5)(130,-17)
\DashLine(130,-17)(150,-25){4}
\Line(0,20)(40,20)
\Line(130,57)(130,-17)
\Line(100,45)(100,-5)
\Line(70,33)(70,7)
\Vertex(130,57){1.5}\Vertex(130,-17){1.5}\Vertex(40,20){1.5}
\Vertex(100,45){1.5}\Vertex(100,-5){1.5}
\Vertex(70,33){1.5}\Vertex(70,7){1.5}
\end{picture}
&
$
\Gamma^2_{[0,0,2]}(q^2)
\rule[-4mm]{0mm}{8mm}
$
&=&
$
(-ig)^7\,
\Gamma^2_{2,0}\Gamma^2_{1,0}\Gamma^2_{0,0}[q^2]^{-3\eps}
$
\\[1.2em]
\hspace{-0.33cm}
\SetScale{0.27}
\SetWidth{2}
\raisebox{0.18cm}{
\begin{picture}(150,10)(0,0)
\DashLine(160,50)(145,44){4}
\Line(145,44)(115,31)
\DashLine(115,31)(40,0){4}
\DashLine(115,-31)(145,-44){4}
\Line(40,0)(115,-31)
\Line(145,-44)(160,-50)
\Line(-7,0)(40,0)
\Line(145,44)(145,-44)
\Line(115,31)(115,-31)
\DashArrowArc(73,-14)(11,-22,158){4}
\DashArrowArc(115,0)(11,270,90){4}
\end{picture}}
&
$
\Gamma^2_{[1,1,1]}(q^2)
\rule[-4mm]{0mm}{8mm}
$
&=&
$
(-ig)^9\Gamma^2_{3,0}\Gamma^2_{1,1}(\Sigma_{0,0})^2[q^2]^{-4\eps}
$
\\[1em]
\hline
\end{tabular*}
\caption{\small Some examples how self-energy and vacuum polarization graphs are built up in Yukawa theory.}
\label{tab:tabelle_eins}
\end{table}
\subsection{QED}

The main difference stem from the presence of a gauge parameter
$\xi$ in the photon propagator and from the $-ie\gamma_{\mu}$
vertex which make the calculations more difficult.  In fact, as we
will see, we will have to deal with matrices for the vertex
corrections~\cite{Delbourgo}.  Nevertheless, the structure for the
translation of a graph to an analytical result will be similar to what we had
in Yukawa theory. Hence we will mainly give the results and just list
the relevant changes.

\subsubsection{Vacuum polarization}
We will only consider one-loop photon self-energies, vacuum
polarizations, as in our restricted class of Feynman graphs we can
not construct a gauge invariant set of vacuum polarizations at
higher loop orders.

\subsubsection{Fermion self-energy}

The fermionic propagator in QED with insertions at the fermionic
and photonic lines demands  $\tilde{\Sigma}$:
\begin{eqnarray}
\tilde{\Sigma}_{i,j}
\label{Tilsig00}
&=&
\frac{1}{2}
\left[\,
(2-D)\,F_{j\eps,1+i\eps}-(3-D)\,F_{1+j\eps,i\eps}-(3-D)\,F_{1+j\eps,1+i\eps}
\right.
\nonumber\\
&&
\left.
+\,F_{2+j\eps,-1+i\eps}-2\,F_{2+j\eps,i\eps}+F_{2+j\eps,1+i\eps}\,
\right](1-\delta_{0,j})\nonumber\\
 & & +\,[\,(2-D)\,\Sigma_{i,0}
+ \xi\; \Sigma_{i,0}^{'}\,]\:\delta_{0,j},
\end{eqnarray}
with
\begin{eqnarray}
\Sigma_{i,0} &=& \frac{1}{2}
\left[F_{1+i\varepsilon,1}+F_{i\varepsilon,1}\right]\\
\Sigma_{i,0}^{'} &=& \frac{1}{2}
\left[F_{-1+i\varepsilon,2}-2F_{i\varepsilon,2}-F_{i\varepsilon,1}+F_{1+i\varepsilon,2}-F_{1+i\varepsilon,1}\right].
\end{eqnarray}
The appearance of the Kronecker $\delta_{0,j}$ is obvious from the
fact that the presence of one-loop vacuum polarizations in the
internal photon line forces transversality of that propagation.

\subsubsection{Vertex corrections}
\label{subsec:vertex_QED}

\begin{figure}[t]
\SetScale{1}
\begin{center}\begin{picture}(220,110)(0,0)
\ArrowLine(150,95)(125,85)
\ArrowLine(125,85)(40,50)
\ArrowLine(40,50)(125,15)
\ArrowLine(125,15)(150,5)
\Photon(0,50)(40,50){4}{5.5}
\Photon(125,85)(125,15){4}{8.5}
\Vertex(125,85){1.5}\Vertex(125,15){1.5}\Vertex(40,50){1.5}
\PText(148,106)(0)[]{q}
\PText(148,0)(0)[]{q}
\PText(85,80)(0)[]{k}
\PText(85,22)(0)[]{k}
\PText(142,50)(0)[]{q-k}
\PText(22,65)(0)[]{0}
\Text(40,62)[]{\(\gamma_{\mu}\)}
\Text(124,96)[]{\(\gamma_{\beta}\)}
\Text(124,4)[]{\(\gamma_{\alpha}\)}
\Text(115,50)[]{$\downarrow$}
\end{picture}\end{center}
\caption{\small One-loop contribution to the vertex correction.}
\label{fig:one_loop_vertex}
\end{figure}
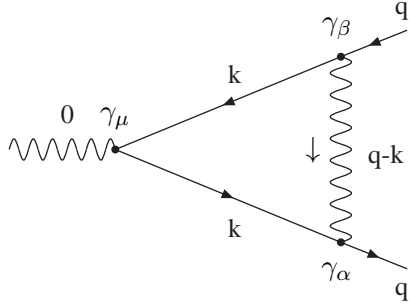

The most important difference to the Yukawa theory occurs in this
part of the calculation.  Consider the one-loop vertex correction
in QED for vertex graphs as shown in
Fig.~\ref{fig:one_loop_vertex}.

The external structure of the vertex consists of two form factors: One
for $\gamma_{\mu}$ and one for
$\frac{q_{\mu}q\hspace{-1.5mm}{/}}{q^2}$, as
$\Gamma_\mu(0,q,q)=F_1(q^2)\gamma_\mu+F_2(q^2)\frac{q_{\mu}q\hspace{-1.5mm}{/}}{q^2}$.
In the previous calculations, every subdivergence just caused an
change in the exponent of various momenta in denominators of the
integrals, resulting in non-integral exponents in equation~(\ref{PasTar}). Now the result
has two form factors $F_1,F_2$ that emerge through the evaluation of a
subgraph in a ``bigger'' graph.  We will administer the two
form factors in a matrix notation in accordance with~\cite{Delbourgo}.
We will need two-by-two matrices. The four entries determine four
functions $\Delta_{a,b}^{(i,j)}$. Here, $a,b\in\{1,2\}$ and $(i,j)$
count the number of internal insertions as before. The case $b=1$
corresponds to an internal vertex $\gamma_\mu$, the case $b=2$
corresponds to an internal vertex $q_\mu q\hspace{-1.8mm}{/}/q^2$
(where $q$, say, is the momentum flowing through this zero-momentum
transfer vertex), while the index $a$ enumerates the two possible
form factors in the result.  The result for the one-loop graph of
Fig.\ref{fig:one_loop_vertex} is then
$$\left[\Delta_{1,1}^{(0,0)}\gamma_\mu+\Delta_{2,1}^{(0,0)}\frac{q_\mu
    q\hspace{-1.8mm}{/}}{q^2}\right][q^2]^{-\eps}.$$

\begin{figure}[b]
\SetScale{1}
\begin{center}\begin{picture}(200,100)(0,0)
\ArrowLine(150,95)(130,87)
\ArrowLine(130,87)(90,71)
\ArrowLine(90,71)(40,50)
\ArrowLine(40,50)(90,29)
\ArrowLine(90,29)(130,13)
\ArrowLine(130,13)(150,5)
\Photon(0,50)(40,50){4}{5.5}
\Photon(130,87)(130,13){4}{7.5}
\Photon(90,71)(90,29){4}{5.5}
\Vertex(130,87){1.5}\Vertex(130,13){1.5}\Vertex(40,50){1.5}
\Vertex(90,71){1.5}\Vertex(90,29){1.5}
\PText(150,108)(0)[]{q}
\PText(150,0)(0)[]{q}
\PText(110,95)(0)[]{k}
\PText(110,11)(0)[]{k}
\PText(65,77)(0)[]{l}
\PText(65,26)(0)[]{l}
\PText(152,50)(0)[]{q-k}
\PText(109,50)(0)[]{k-l}
\PText(20,63)(0)[]{0}
\Text(140,50)[]{$\downarrow$}
\Text(101,50)[]{$\downarrow$}
\Text(90,19)[]{\(\gamma_{\sigma}\)}
\Text(40,60)[]{\(\gamma_{\mu}\)}
\Text(90,81)[]{\(\gamma_{\nu}\)}
\Text(130,3)[]{\(\gamma_{\alpha}\)}
\Text(130,97)[]{\(\gamma_{\beta}\)}
\end{picture}\end{center}
\vspace{0.2cm}
\caption{\small Two-loop contribution to the vertex correction.}
\label{fig:two_loop_vertex}
\end{figure}
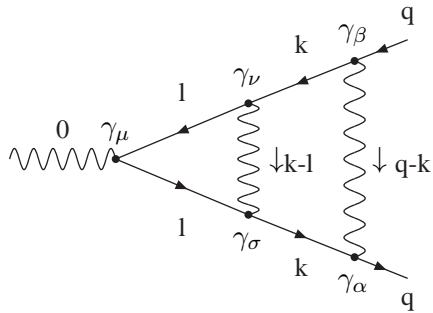
\noindent For the two-loop graph shown in figure~\ref{fig:two_loop_vertex} we find
\begin{eqnarray}
\Gamma_{[0,0,1]} = \left[ \left(
\Delta^{(0,0)}_{11}\Delta^{(1,0)}_{11}
+\Delta^{(0,0)}_{21}\Delta^{(1,0)}_{12} \right) \gamma_{\mu} +
\left( \Delta^{(0,0)}_{11}\Delta^{(1,0)}_{21}
+\Delta^{(0,0)}_{21}\Delta^{(1,0)}_{22} \right)
\frac{q_{\mu}q\hspace{-1.8mm}{/}}{q^2} \right]
[q^2]^{-2\eps}.&&\label{g2p2}
\end{eqnarray}
The multiplication of the $\Delta^{(i,j)}_{kl}$ can be
reformulated as a matrix multiplication. We define the general
matrix (the upper index refers to the two cases of vertex
corrections which we consider, we omit that index for simplicity
in the matrix entries):
\begin{eqnarray*}
M^1_{i,j}
:=
\left(
\begin{array}{cc}
\Delta_{11}^{(i,j)}&\Delta_{12}^{(i,j)}\\
\Delta_{21}^{(i,j)}&\Delta_{22}^{(i,j)}
\end{array}
\right)
\end{eqnarray*}
To each part of a graph which has the form of a vertex correction we
assign a matrix. In this case of vertex corrections of type one, as
in~\ref{fig:two_vertex_types}.a, the index i is the total number of
subdivergences at the fermionic line (with no difference if it is of
the form $\Sigma$ or $\Gamma$) and j the number of subdivergences at
the photon line, as in Yukawa theory. One multiplies the matrices for
the different divergent parts of a graph, starting from the innermost
vertex correction which contains the vertex of zmt, and the
corresponding matrix has zero entries in the second column obviously.

Let us make this clear with the help of our example of the
two-loop vertex correction again. For the graph of
figure~\ref{fig:two_loop_vertex} this means that we begin with
the inner vertex correction marked with a box:
\begin{center}
\begin{tabular}{cc}
\SetScale{0.70}
\begin{picture}(150,30)(0,0)
\ArrowLine(170,45)(150,37)
\ArrowLine(150,37)(110,21)
\ArrowLine(110,21)(60,0)
\ArrowLine(60,0)(110,-21)
\ArrowLine(110,-21)(150,-37)
\ArrowLine(150,-37)(170,-45)
\Photon(20,0)(60,0){4}{5.5}
\Photon(150,37)(150,-37){4}{7.5}
\Photon(110,21)(110,-21){4}{5.5}
\Vertex(150,37){1.5}\Vertex(150,-37){1.5}\Vertex(60,0){1.5}
\Vertex(110,21){1.5}\Vertex(110,-21){1.5}
\Line(120,-28)(120,28)
\Line(53,-28)(120,-28)
\Line(53,-28)(53,28)
\Line(53,28)(120,28)
\end{picture}
\SetScale{1}
&
$
\label{M0}
\equiv
M^1_{0,0}
=
\left(
\begin{array}{cc}
\Delta_{11}^{(0,0)}&  0\\
\Delta_{21}^{(0,0)}& 0
\end{array}
\right)
$\\
&\\
&\\
\end{tabular}
\end{center}
and then multiply this with the matrix for the outer vertex
correction, which has one vertex correction as a subdivergence:
\begin{center}
\begin{tabular}{cc}
\SetScale{0.70}
\begin{picture}(150,30)(0,0)
\ArrowLine(170,45)(150,37)
\ArrowLine(150,37)(110,21)
\ArrowLine(110,21)(60,0)
\ArrowLine(60,0)(110,-21)
\ArrowLine(110,-21)(150,-37)
\ArrowLine(150,-37)(170,-45)
\Photon(20,0)(60,0){4}{5.5}
\Photon(150,37)(150,-37){4}{7.5}
\Photon(110,21)(110,-21){4}{5.5}
\Vertex(150,37){1.5}\Vertex(150,-37){1.5}\Vertex(60,0){1.5}
\Vertex(110,21){1.5}\Vertex(110,-21){1.5}
\Line(160,-44)(160,44)
\Line(53,-44)(160,-44)
\Line(53,-44)(53,44)
\Line(53,44)(160,44)
\end{picture}
\SetScale{1}
&
$
\equiv
M^1_{1,0}
=
\left(
\begin{array}{cc}
\Delta_{11}^{(1,0)}&\Delta_{12}^{(1,0)}\\
\Delta_{21}^{(1,0)}&\Delta_{22}^{(1,0)}
\end{array}
\right)
$
\\&\\&\\
\end{tabular}
\end{center}
\begin{eqnarray}
\Rightarrow\quad\quad
\left(
\begin{array}{cc}
\Delta_{11}^{(1,0)}&\Delta_{12}^{(1,0)}\\
\Delta_{21}^{(1,0)}&\Delta_{22}^{(1,0)}
\end{array}
\right)
\left(
\begin{array}{cc}
\Delta_{11}^{(0,0)}& 0\\ \Delta_{21}^{(0,0)}& 0
\end{array}
\right)
=
\left(
\begin{array}{cc}
\label{matrix_result}
\Delta_{11}^{(1,0)}\Delta_{11}^{(0,0)}+\Delta_{12}^{(1,0)}\Delta_{21}^{(0,0)}& 0\\
\Delta_{21}^{(1,0)}\Delta_{11}^{(0,0)}+\Delta_{22}^{(1,0)}\Delta_{21}^{(0,0)}&
0
\end{array}
\right).
\end{eqnarray}
Multiplying (\ref{matrix_result}) with $(1,0)^T[q^2]^{-2\eps}$,
we get the result as the column vector
\begin{eqnarray*}
\left(
\begin{array}{c}
(\Delta_{11}^{(1,0)}\Delta_{11}^{(0,0)}+\Delta_{12}^{(1,0)}\Delta_{21}^{(0,0)})[q^2]^{-2\eps}\\
(\Delta_{21}^{(1,0)}\Delta_{11}^{(0,0)}+\Delta_{22}^{(1,0)}\Delta_{21}^{(0,0)})[q^2]^{-2\eps}
\end{array}
\right),
\end{eqnarray*}
where in general the ``upper'' entry of this vector is the form factor
$F_1(q^2)$ belonging to $\gamma_\mu$ and the ``lower'' one the form factor
$F_2(q^2)$ belonging to $\frac{q^{\mu} q\hspace{-1.5mm}{/}}{q^2}$.

Subdivergences that are {\em not} vertex corrections given by
$\Sigma_{i,j}$ and $\Pi_{i,j}$ are multiplied with a unit matrix
and inserted in the string of matrices in front of the matrix of
the vertex correction which they are part of. They increase
subscripts $i,j$ accordingly in that matrix.

As an example we get for the following graph:
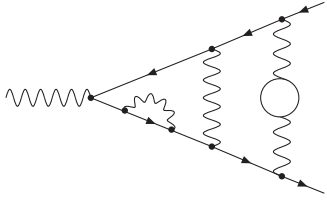
\begin{center}
\SetScale{0.8}
\begin{picture}(300,80)(0,0)
\ArrowLine(170,95)(150,87)
\ArrowLine(150,87)(117,73)
\ArrowLine(117,73)(60,50)
\ArrowLine(60,50)(117,27)
\ArrowLine(117,27)(150,13)
\ArrowLine(150,13)(170,5)
\Photon(20,50)(60,50){4}{5.5}
\Photon(150,87)(150,13){4}{8}
\Photon(117,73)(117,27){4}{5.5}
\Vertex(150,87){1.5}\Vertex(150,13){1.5}\Vertex(60,50){1.5}
\Vertex(117,73){1.5}\Vertex(117,27){1.5}
\Vertex(76,44){1.5}\Vertex(98,35){1.5}
\PhotonArc(87,39)(11,-22,158){2}{5.5}
\BCirc(149,50){9}
\end{picture}
\SetScale{1}
\end{center}
\begin{eqnarray*}
\label{mpms}
&\equiv&
\left(
\begin{array}{cc}
\Delta_{11}^{(2,1)}&\Delta_{12}^{(2,1)}\\
\Delta_{21}^{(2,1)}&\Delta_{22}^{(2,1)}
\end{array}
\right)
\left(
\begin{array}{cc}
\Pi_{0,0}&0\\
0&\Pi_{0,0}
\end{array}
\right)
\left(
\begin{array}{cc}
\Delta_{11}^{(1,0)}&\Delta_{12}^{(1,0)}\\
\Delta_{21}^{(1,0)}&\Delta_{22}^{(1,0)}
\end{array}
\right)
\left(
\begin{array}{cc}
\tilde{\Sigma}_{0,0}&0\\
0&\tilde{\Sigma}_{0,0}
\end{array}
\right)\\
\nonumber\\
&\equiv&
\label{mpms2}
M^1_{2,1}\Pi_{0,0}M^1_{1,0}\tilde{\Sigma}_{0,0}.
\end{eqnarray*}

As already mentioned, the indices $i$ and $j$ count for QED vertex
corrections of type one, $\Gamma^1_{[i,j,k]}$, similar to the ones
defined for Yukawa vertex corrections of this type. Similarly, the
indices for the type two vertex corrections $\Gamma^2_{[i,j,k]}$
in QED count as in Yukawa type two. And the matrix multiplication
described above stays the same apart from different entries for
the matrices, which can be readily calculated in terms of
functions $F_{a,b}$. We do not list the two sets of four functions
$\Delta^{(i,j)}_{a,b}$ in terms of $F_{c,d}$ functions explicitly,
but the interested reader can find them from our publicly
available programs~\cite{Isaurl}. Table~\ref{tab:tabelle_zwei}
shows some examples how graphs are built in QED.

\begin{center}
\hspace{-2cm}
\begin{table}[t]
\begin{tabular*}{\textwidth}{@{}l@{\extracolsep{\fill}}l@{\extracolsep{1ex}}l@{\extracolsep{1ex}}l@{}}\hline
\SetScale{0.3}
\SetWidth{1.5}
\begin{picture}(55,23)(0,0)
\Line(0,0)(160,0)
\PhotonArc(80,0)(40,0,180){4}{9.5}
\Vertex(120,0){1.5}\Vertex(40,0){1.5}
\end{picture}
\SetScale{1}
&
$
\tilde{\Sigma}_{[0,0]}(q^2)
\rule[-4mm]{0mm}{8mm}
$
&
=
&
$
(-ie)^2\,
\tilde{\Sigma}_{0,0}[q^2]^{-\eps}
q\hspace{-2mm}{/}
$
\\
\SetScale{0.3}
\SetWidth{1.5}
\begin{picture}(55,23)(1,0)
\Line(0,0)(160,0)
\PhotonArc(80,0)(60,0,180){4}{12.5}
\PhotonArc(80,0)(25,0,180){4}{6.5}
\Vertex(140,0){1.5}\Vertex(20,0){1.5}
\Vertex(105,0){1.5}\Vertex(55,0){1.5}
\end{picture}
&
$
\tilde{\Sigma}_{[1,0]}(q^2)
\rule[-4mm]{0mm}{8mm}$
&=&
$
(-ie)^4
\tilde{\Sigma}_{1,0}
\tilde{\Sigma}_{0,0}
[q^2]^{-2\eps}
q\hspace{-2mm}{/}
$\\
\SetScale{0.16}
\SetWidth{2}
\begin{picture}(55,23)(-2,0)
\Line(-20,0)(30,0)
\Line(30,0)(60,0)
\Line(60,0)(110,0)
\Line(110,0)(150,0)
\Line(150,0)(200,0)
\Line(200,0)(230,0)
\Line(230,0)(280,0)
\PhotonArc(130,0)(100,0,180){4}{17.5}
\PhotonArc(85,0)(25,0,180){4}{6.5}
\PhotonArc(175,0)(25,0,180){4}{6.5}
\end{picture}
&
$
\tilde{\Sigma}_{[2,0]}(q^2)
\rule[-4mm]{0mm}{8mm}
$
&=&
$
(-ie)^6
\tilde{\Sigma}_{2,0}\tilde{\Sigma}_{0,0}\tilde{\Sigma}_{0,0}
[q^2]^{-3\eps}q\hspace{-2mm}{/}
$
\\
\SetScale{0.16}
\SetWidth{2}
\begin{picture}(55,23)(-1.5,0)
\Line(-20,0)(30,0)
\Line(30,0)(60,0)
\Line(60,0)(110,0)
\Line(110,0)(150,0)
\Line(150,0)(200,0)
\Line(200,0)(230,0)
\Line(230,0)(280,0)
\PhotonArc(130,0)(100,0,180){4}{17.5}
\PhotonArc(85,0)(25,0,180){4}{6.5}
\PhotonArc(175,0)(25,0,180){4}{6.5}
\Vertex(30,0){1.5}\Vertex(230,0){1.5}
\Vertex(60,0){1.5}\Vertex(110,0){1.5}
\Vertex(150,0){1.5}\Vertex(200,0){1.5}
\BCirc(130,100){22}
\Vertex(108,100){1.5}\Vertex(152,100){1.5}
\end{picture}
&
$
\tilde{\Sigma}_{[2,1]}(q^2)
\rule[-4mm]{0mm}{8mm}
$
&=&
$
(-ie)^8
\tilde{\Sigma}_{2,1}
\Pi_{0,0}
(\tilde{\Sigma}_{0,0})^2
[q^2]^{-4\eps}
q\hspace{-2mm}{/}
$
\\
\SetScale{0.3}
\SetWidth{1.5}
\begin{picture}(55,23)(1.5,0)
\Photon(0,20)(45,20){4}{6.5}
\Photon(115,20)(160,20){4}{6.5}
\BCirc(80,20){35}
\end{picture}
&
$
\Pi_{[0,0]}(q^2)
\rule[-4mm]{0mm}{8mm}
$&=&
$
(-ie)^2
[q^2]^{-\eps}
\Pi_{0,0}
\left[g^{\mu\nu}-\frac{q^{\mu}q^{\nu}}{q^2}\right]
q^2
$
\\
\SetScale{0.3}
\SetWidth{1.5}
\begin{picture}(55,23)(0,0)
\Line(150,65)(125,55)
\Line(125,55)(40,20)
\Line(40,20)(125,-15)
\Line(125,-15)(150,-25)
\Photon(0,20)(40,20){4}{5.5}
\Photon(125,55)(125,-15){4}{8.5}
\Vertex(125,55){1.5}\Vertex(125,-15){1.5}\Vertex(40,20){1.5}
\end{picture}
&
$
\Gamma^1_{[0,0,0]}(q^2)
\rule[-4mm]{0mm}{8mm}
$&=&
$
(-ie)^3\,
M^1_{0,0}[q^2]^{-\eps}
$
\\
\SetScale{0.3}
\SetWidth{1.5}
\begin{picture}(55,23)(0,0)
\Line(150,65)(130,57)
\Line(130,57)(100,45)
\Line(100,45)(70,33)
\Line(70,33)(40,20)
\Line(40,20)(70,7)
\Line(70,7)(100,-5)
\Line(100,-5)(130,-17)
\Line(130,-17)(150,-25)
\Photon(0,20)(40,20){4}{5.5}
\Photon(130,57)(130,-17){4}{7.5}
\Photon(100,45)(100,-5){4}{5.5}
\Photon(70,33)(70,7){4}{3.5}
\Vertex(130,57){1.5}\Vertex(130,-17){1.5}\Vertex(40,20){1.5}
\Vertex(100,45){1.5}\Vertex(100,-5){1.5}
\Vertex(70,33){1.5}\Vertex(70,7){1.5}
\end{picture}
&
$
\Gamma^1_{[0,0,2]}(q^2)
\rule[-4mm]{0mm}{8mm}
$&=&
$
(-ie)^7\,
M^1_{2,0}M^1_{1,0}M^1_{0,0}[q^2]^{-3\eps}
$
\\
\SetScale{0.3}
\SetWidth{1.5}
\begin{picture}(55,23)(0,0)
\Line(150,65)(125,55)
\Line(125,55)(40,20)
\Line(40,20)(125,-15)
\Line(125,-15)(150,-25)
\Photon(0,20)(40,20){4}{5.5}
\Photon(125,55)(125,-15){4}{8.5}
\Vertex(125,55){1.5}\Vertex(125,-15){1.5}\Vertex(40,20){1.5}
\PhotonArc(82,3)(22,-22,158){2}{8.5}
\PhotonArc(82,3)(10,-22,158){2}{6.5}
\end{picture}
&
$
\Gamma^1_{[2,0,0]}(q^2)
\rule[-4mm]{0mm}{8mm}
$&=&
$
(-ie)^7\,
M^1_{2,0}\tilde{\Sigma}_{1,0}\tilde{\Sigma}_{0,0}[q^2]^{-3\eps}
$
\\[.5em]
\SetScale{0.3}
\SetWidth{1.5}
\begin{picture}(55,23)(0,0)
\Line(150,65)(125,55)
\Line(125,55)(40,20)
\Line(40,20)(125,-15)
\Line(125,-15)(150,-25)
\Photon(0,20)(40,20){4}{5.5}
\Photon(125,55)(125,-15){4}{8.5}
\Vertex(125,55){1.5}\Vertex(125,-15){1.5}\Vertex(40,20){1.5}
\BCirc(125,20){14}
\end{picture}
&
$
\Gamma^1_{[0,1,0]}(q^2)
\rule[-4mm]{0mm}{8mm}
$&=&
$
(-ie)^5\,
M^1_{0,1}\Pi_{0,0}[q^2]^{-2\eps}
$
\\[.5em]
\SetScale{0.3}
\SetWidth{2}
\begin{picture}(55,23)(0,0)
\Line(150,65)(125,55)
\Photon(125,55)(40,20){4}{5.5}
\Line(40,20)(125,-15)
\Photon(125,-15)(150,-25){4}{2.5}
\Line(0,20)(40,20)
\Line(125,55)(125,-15)
\Vertex(125,55){1.5}\Vertex(125,-15){1.5}\Vertex(40,20){1.5}
\end{picture}
&
$
\Gamma^2_{[0,0,0]}(q^2)
\rule[-4mm]{0mm}{8mm}
$
&=&$
(-ig)^3\,
M^2_{0,0}[q^2]^{-\eps}
$
\\[.0em]
\hspace{0.001cm}
\SetScale{0.3}
\SetWidth{2}
\begin{picture}(55,23)(3,0)
\Line(150,65)(130,57)
\Photon(130,57)(100,45){4}{3}
\Line(100,45)(70,33)
\Photon(70,33)(40,20){4}{3}
\Line(40,20)(70,7)
\Photon(70,7)(100,-5){4}{3}
\Line(100,-5)(130,-17)
\Photon(130,-17)(150,-25){4}{2.5}
\Line(0,20)(40,20)
\Line(130,57)(130,-17)
\Line(100,45)(100,-5)
\Line(70,33)(70,7)
\Vertex(130,57){1.5}\Vertex(130,-17){1.5}\Vertex(40,20){1.5}
\Vertex(100,45){1.5}\Vertex(100,-5){1.5}
\Vertex(70,33){1.5}\Vertex(70,7){1.5}
\end{picture}
&
$
\Gamma^2_{[0,0,2]}(q^2)
\rule[-4mm]{0mm}{8mm}
$
&=&
$
(-ig)^7\,
M^2_{2,0}M^2_{1,0}M^2_{0,0}[q^2]^{-3\eps}
$
\\[1.2em]
\hspace{-0.2cm}
\SetScale{0.27}
\SetWidth{2}
\raisebox{0.18cm}{
\begin{picture}(150,10)(0,0)
\Photon(160,50)(145,44){4}{1.5}
\Line(145,44)(115,31)
\Photon(115,31)(40,0){4}{5.5}
\Photon(115,-31)(145,-44){4}{2.5}
\Line(40,0)(115,-31)
\Line(145,-44)(160,-50)
\Line(-7,0)(40,0)
\Line(145,44)(145,-44)
\Line(115,31)(115,-31)
\PhotonArc(73,-14)(11,-22,158){4}{3.5}
\PhotonArc(115,0)(11,270,90){4}{3.5}
\end{picture}}
&
$
\Gamma^2_{[1,1,1]}(q^2)
\rule[-4mm]{0mm}{8mm}
$
&=&
$
(-ie)^{9}
M^2_{3,0}M^2_{1,1}(\tilde{\Sigma}_{0,0})^2[q^2]^{-4\eps}
$
\\[1em]
\hline
\end{tabular*}
\caption{\small Some examples how self-energy and vacuum polarization graphs are built up in QED.}
\label{tab:tabelle_zwei}
\end{table}
\end{center}

\section{Renormalization}
Renormalization employs a simple principle of multiplicative
subtraction, making use of the underlying Hopf algebra structure
of Feynman graphs~\cite{Hopfalgebra,RHII}: the coproduct
$$\Delta(\Gamma)=\Gamma\otimes
1+1\otimes\Gamma+\sum_{\gamma\subset\Gamma}\gamma\otimes\Gamma/\gamma$$
and antipode $$
S(\Gamma)=-\Gamma-\sum_{\gamma\subset\Gamma}\gamma\,\Gamma/\gamma$$
are the structure maps which allow the construction of
counterterms and renormalized quantities. One employs Feynman
rules $\phi:H\to V$ as an element in the group of characters of
the Hopf algebra $H$, with target space $V$ (a suitable ring or
algebra) and makes the target space into a Baxter algebra $(V,R)$
by choosing a renormalization map $R$ such  that
$R(ab)+R(a)R(b)=R(aR(b))+R(R(a)b)$. One then has the counterterm
$$S_R^\phi(\Gamma)=-R[\phi(\Gamma)+\sum_{\gamma\subset\Gamma}S_R^\phi(\gamma)\phi(\Gamma/\gamma)]$$
and a further recursion delivers the renormalized result
$$
S_R^\phi\star\phi(\Gamma)=S_R^\phi(\Gamma)+\phi(\Gamma)+\sum_{\gamma\subset\Gamma}S_R^\phi(\gamma)\phi(\Gamma/\gamma).$$
The counterterm $S_R^\phi$ is in the image of $R$, while
$\phi(\Gamma)+\sum_{\gamma\subset\Gamma}S_R^\phi(\gamma)\phi(\Gamma/\gamma)$
is the result of Bogoliubov's famous $\overline{R}(\Gamma)$ operation
on $\Gamma$ which eliminates subdivergences in $\Gamma$~\cite{Collins}.

Under suitable conditions on the behavior of $R$, in this ratio
of characters short distance singularities drop
out~\cite{Hopfalgebra,Connes}.

Before we comment on the renormalization scheme chosen for our calculations,
let us introduce the bidegree of a Feynman graph. This standard notion
can be introduced for any
Hopf algebra which is reduced to scalars \cite{Bigrad,RHII} by the counit
$\bar{e}$ with
$$\bar{e}(q1)=q,\;\bar{e}(X)=0,\;\mbox{else.}$$
If we decompose $H=H_0\oplus H_{\rm aug}$, with $H_{\rm aug}$ being the
augmentation ideal as the kernel
of $\bar{e}$  we can investigate, for any positive integer $k$
and Hopf algebra element $X$,
$$X_k:=\Delta^{k-1}(X)\cap H_{\rm aug}^{\otimes k}.$$
For sufficiently large $k$ this will necessarily vanish.
We define the bidegree ${\rm bid}(X)$ as the largest $k$
such that $X_k\not= 0$. Elements in $H_0$ have bidegree zero.
Note that Hopf algebra elements of unit bidegree are precisely the primitive
elements $X\in H_{\rm aug}$ in the Hopf algebra,
with $$\Delta(X)=X\otimes 1+1\otimes X\not\in H_{\rm aug}\otimes H_{\rm aug}.$$

Having introduced this standard notion we introduce the renormalization
scheme for which we choose minimal subtraction
(MS). Each application of the scheme is given by a projection onto
the pole-part of the considered Laurent series and is symbolized
with brackets ``$\langle \rangle$'':
$$\langle\sum_{j:=-r}^{+\infty}c_j\eps^j\rangle=\sum_{j:=-r}^{-1}c_j\eps^j.$$
The reader should convince himself that this map makes the ring of
Laurent series with poles of finite order into a Baxter algebra,
$$\langle ab\rangle+\langle a\rangle\langle b\rangle=\langle a \langle b \rangle\rangle+
\langle \langle a\rangle b\rangle.$$
Note that the degree $r$ of the pole terms in a Laurent series assigned
to a graph by the Feynman rules in dimensional regularization is in general
majorized by the bidegree
$r\leq {\rm bid}(\Gamma)$ and equals the bidegree in our simple applications.

 We expect to encounter
$\zeta(n)$ inside the coefficients of the Laurent series in the
regularization parameter $\eps$ emerging from a series expansion
of the functions $F_{a,b}$.

In QED we have to take our matrix-calculus into account. The
renormalization of such matrix expressions is now given by
inserting a diagonal matrix $R$, which consists of the
renormalization map as entries:
\begin{eqnarray*}
R:=
\left(
\begin{array}{cc}
R_{MS}&0\\
0&R_{MS}
\end{array}
\right)
\end{eqnarray*}
Inside a string of matrices, this matrix has to be inserted
wherever the renormalization map is applied. It acts on
expressions on the right.
\section{Rooted Trees}\label{sec:antipode}

Before we build up graphs and calculate their counterterms, ie.\
their antipodes, let us first mention that in our simplified
context, the Hopf algebra of Feynman graphs is isomorphic to a
Hopf algebra of rooted trees with a very small set of decorations
given by our one-loop graphs.

\begin{figure}[t]
\vspace{-4em}\hspace{1cm}
\begin{tabular}{ccccc}
\hspace{-2cm}
\SetScale{0.6}
\begin{picture}(40,80)(0,0)
\ArrowLine(180,40)(150,29)
\ArrowLine(150,29)(70,0)
\ArrowLine(70,0)(150,-29)
\ArrowLine(150,-29)(180,-40)
\Photon(30,0)(70,0){4}{6.5}
\Photon(150,29)(150,-29){4}{6.5}
\Vertex(150,29){1.5}\Vertex(150,-29){1.5}\Vertex(70,0){1.5}
\PhotonArc(94,-9)(9,-22,158){2}{5.5}
\PhotonArc(126,-20)(9,-22,158){2}{5.5}
\Line(82,-14)(82,6)
\Line(82,-14)(108,-14)
\Line(108,-14)(108,6)
\Line(82,6)(108,6)
\Line(114,-25)(114,-5)
\Line(114,-25)(140,-25)
\Line(140,-25)(140,-5)
\Line(114,-5)(140,-5)
\Line(65,-34)(163,-34)
\Line(65,-34)(65,34)
\Line(65,34)(163,34)
\Line(163,-34)(163,34)
\Vertex(95,6){5}
\Vertex(127,-5){5}
\Vertex(112,34){5}
\Line(95,6)(112,34)
\Line(127,-5)(112,34)
\end{picture}
&
\hspace{1cm}
$\Rightarrow$
\hspace{1.5cm}
&
\hspace{-1.5cm}
\SetScale{0.9}
\begin{picture}(70,70)(0,0)
\Vertex(10,-5){5}
\Vertex(40,-5){5}
\Vertex(25,15){5}
\Line(10,-5)(25,15)
\Line(25,15)(40,-5)
\SetScale{0.2}
\Line(-110,-43)(-10,-43)
\PhotonArc(-60,-43)(40,0,180){4}{9.5}
\SetScale{1}
\SetScale{0.2}
\Line(250,-43)(350,-43)
\PhotonArc(300,-43)(40,0,180){4}{9.5}
\ArrowLine(365,125)(335,114)
\ArrowLine(335,114)(255,85)
\ArrowLine(255,85)(335,56)
\ArrowLine(335,56)(365,45)
\Photon(220,85)(255,85){4}{6.5}
\Photon(335,114)(335,56){4}{6.5}
\SetScale{1}
\end{picture}
&
$\doteq$
&
((\(\Sigma\))(\(\Sigma\))\(\Gamma\))
\\[-.5em]
\hspace{-2cm}
\SetScale{0.6}
\begin{picture}(40,100)(0,0)
\ArrowLine(180,40)(150,29)
\ArrowLine(150,29)(70,0)
\ArrowLine(70,0)(150,-29)
\ArrowLine(150,-29)(180,-40)
\Photon(30,0)(70,0){4}{6.5}
\Photon(150,29)(150,-29){4}{6.5}
\Vertex(150,29){1.5}\Vertex(150,-29){1.5}\Vertex(70,0){1.5}
\PhotonArc(109,-14)(11,-22,158){2}{5.5}
\BCirc(151,0){9}
\Line(96,-20)(96,0)
\Line(96,-20)(123,-20)
\Line(123,-20)(123,0)
\Line(96,0)(123,0)
\Line(162,-11)(162,11)
\Line(162,11)(140,11)
\Line(140,-11)(140,11)
\Line(162,-11)(140,-11)
\Line(65,-34)(165,-34)
\Line(65,-34)(65,40)
\Line(65,40)(165,40)
\Line(165,-34)(165,40)
\Vertex(109,0){5}
\Vertex(151,11){5}
\Vertex(115,40){5}
\Line(109,0)(115,40)
\Line(151,11)(115,40)
\end{picture}
\SetScale{1}
&
\hspace{1cm}
$\Rightarrow$
\hspace{1.5cm}
&
\hspace{-1.5cm}
\SetScale{0.9}
\begin{picture}(70,70)(0,0)
\Vertex(10,-5){5}
\Vertex(40,-5){5}
\Vertex(25,15){5}
\Line(10,-5)(25,15)
\Line(25,15)(40,-5)
\SetScale{0.2}
\Line(-110,-43)(-10,-43)
\PhotonArc(-60,-43)(40,0,180){4}{9.5}
\SetScale{1}
\SetScale{0.5}
\BCirc(130,-11){12}
\Photon(100,-11)(119,-11){4}{3}
\Photon(141,-11)(160,-11){4}{3}
\SetScale{0.2}
\ArrowLine(365,125)(335,114)
\ArrowLine(335,114)(255,85)
\ArrowLine(255,85)(335,56)
\ArrowLine(335,56)(365,45)
\Photon(220,85)(255,85){4}{6.5}
\Photon(335,114)(335,56){4}{6.5}
\SetScale{1}
\end{picture}
&
$\doteq$
&
((\(\Sigma\))(\(\Pi\))\(\Gamma\))
\\[-.5em]
\SetScale{0.30}
\begin{picture}(60,100)(0,0)
\PhotonArc(150,-30)(110,0,180){4}{17.5}
\PhotonArc(150,-30)(60,0,180){4}{13}
\PhotonArc(150,-30)(15,0,180){4}{5.5}
\Line(32,-50)(268,-50)
\Line(10,-30)(290,-30)
\Line(32,-50)(32,90)
\Line(268,-50)(268,90)
\Line(32,90)(268,90)
\Line(128,-40)(172,-40)
\Line(128,-40)(128,-5)
\Line(172,-40)(172,-5)
\Line(128,-5)(172,-5)
\Line(83,-45)(217,-45)
\Line(83,40)(217,40)
\Line(83,-45)(83,40)
\Line(217,-45)(217,40)
\Vertex(150,-5){10}
\Vertex(150,40){10}
\Vertex(150,90){10}
\Line(150,40)(150,-5)
\Line(150,90)(150,30)
\end{picture}
&
\hspace{1cm}
$\Rightarrow$
\hspace{1.5cm}
&
\hspace{-0.5cm}
\SetScale{0.9}
\begin{picture}(70,70)(0,0)
\Vertex(0,-10){5}
\Vertex(0,10){5}
\Vertex(0,30){5}
\Line(0,-10)(0,10)
\Line(0,10)(0,30)
\SetScale{0.2}
\Line(100,35)(200,35)
\PhotonArc(150,35)(40,0,180){4}{9.5}
\Line(100,135)(200,135)
\PhotonArc(150,135)(40,0,180){4}{9.5}
\Line(100,-66)(200,-66)
\PhotonArc(150,-66)(40,0,180){4}{9.5}
\end{picture}
&
$\doteq$
&
(((\(\Sigma\))\(\Sigma\))\(\Sigma\))\\[2em]
\end{tabular}
\caption{\small Translation of Feynman diagrams into rooted trees.
  Here we are dealing with decorated rooted trees, where the
  subdivergences are written next to the vertex to which they belong.
   In the last
  column one can see another way to encode the structure of the trees
  in the form of nested lists~\cite{Hopfalgebra}: The entries are again the
  divergent parts of a Feynman diagram and the formation of the
  parentheses gives the structure of the tree, beginning with the
  vertices ``at the end'' of a tree up to the root.}
\label{fig:translation}
\end{figure}
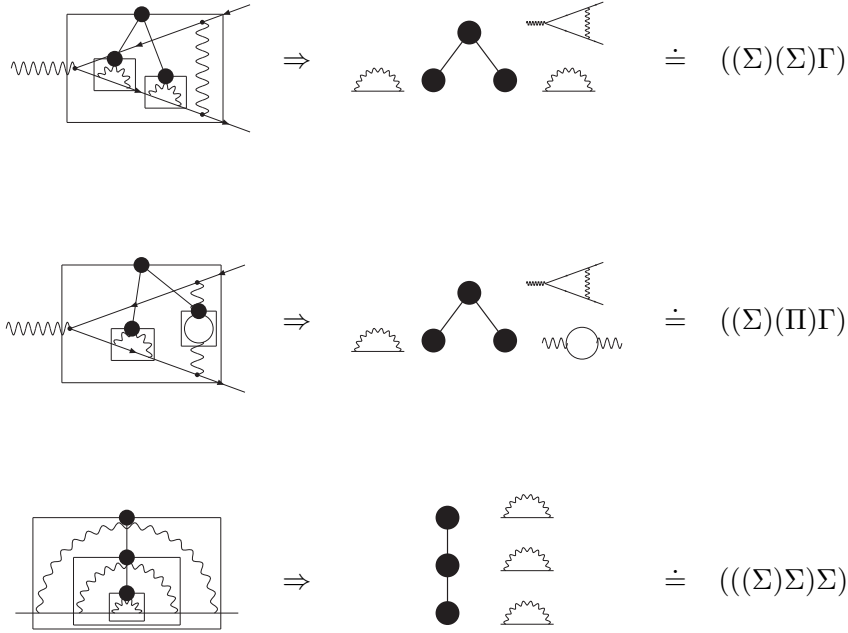

In the class of graphs to which we have restricted ourselves
one-particle irreducible subgraphs are either nested in each other, or
disjoint. They hence form tree-like hierarchies, and one easily
translates graphs in rooted trees~\cite{Connes}, with a one-to-one
correspondence between one-particle irreducible subgraphs and vertices
in the rooted tree (the map from graphs to rooted trees is one-to-many
for overlapping divergent graphs, and can be systematically
constructed~\cite{Overlapping}):

The translation from a Feynman diagram to a rooted tree has to be done
in the following way: {\sl Set a box around the subdivergences of a
  Feynman graph $\Gamma$ and mark the upper horizontal line with a dot
  ($\simeq$\,vertex).  Dots of nested boxes, that is boxes where one
  of them is contained inside the other, are connected with a line
  ($\simeq$\,edge)} (see fig.~\ref{fig:translation}).

Each vertex of such a tree represents a divergent subgraph of the
diagram. The antipode of such a tree is a sum over all {\em full
cuts} or, recursively, iterates over admissible cuts using the
coproduct. This easy-to-implement~\cite{BrKr} coproduct
$$\Delta(T)=T\otimes 1+1\otimes T+\sum_{{\mbox{\tiny admissible
cuts $c$}} }P^c(T)\otimes R^c(T)$$ is by now a standard
combinatorial tool (see, for example~\cite{Connes}). Using it, the
counterterm is a recursive construct
$$
S_{MS}(T)=-\langle T(\Gamma)+\sum_{\mbox{\tiny admissible cuts
$c$}}S_{MS}(P^c(T)) R^c(T)\rangle,$$ where an evaluation of the graph
corresponding to the tree using the Feynman rules is understood
before applying the projection $\langle \rangle$ for minimal subtraction (MS),
in accordance with the general formula given for graphs.


\section{Implementation}\label{sec:implementation}

We shortly describe an implementation of our scheme in a computer
program.  This implementation is not entirely self-contained in the
sense that given a Feynman diagram, it would compute the antipode.
Instead, its input is a decorated rooted tree in list notation such as
\((\Gamma(\Pi)(\Sigma))\). The actual construction of that rooted tree
from the diagram, as described in section \ref{sec:antipode} is left
to the user (see also fig.~\ref{fig:translation}).

Our implementation uses the GiNaC system for symbolic computation in
the C++ programming language~\cite{GiNaCpaper}.  GiNaC provides
efficient implementations for handling Laurent series as needed in
dimensional regularization.  From C++, all the linguistic instruments
for object-oriented programming are borrowed and available to us.  We
follow a traditional approach for representation of our rooted trees
where there exists a container class called {\tt node} that may or may
not have several children of the same class.  In addition, each vertex
in an object of class {\tt node} contains an object of the abstract
and polymorphic class {\tt decoration}.  From the abstract {\tt
  decoration} class a number of concrete classes like {\tt Sigma},
{\tt Gamma} and {\tt Vacuum} are derived, corresponding to primitive
Feynman diagrams \(\Sigma\), \(\Gamma\) and \(\Pi\), as well as some
additional classes that allow us to distinguish between the position
of a subgraph inside its supergraph.

The trees under consideration are not ordered with respect to their
children, i.e.\ the trees \((\Gamma(\Sigma)(\Pi))\) and
\((\Gamma(\Pi)(\Sigma))\) are equivalent.  Therefore the children form
a multiset where only the multiplicity of occurrences is relevant.

A template class {\tt multiset} is part of the C++ Standard Template
Library (STL).  It has the additional advantage that the elements are
always automatically sorted with respect to some specified ordering.
This turns out to be useful for convenient identification of
equivalent nodes and also to establish an order relation on them.  The
state of the edges leading to the children of a node (either ``cut''
or ``uncut'') needs to be taken into account as well, so the multiset
is really one of pairs of nodes and boolean variables. We chose to
delegate methods from class {\tt node} to the corresponding class
derived from {\tt decoration} using dynamic type-dispatch.  Hence, the
decoration must be stored as a pointer, calling for some hand-made
memory management in class {\tt node}.  A completely realistic layout
of our class {\tt node} is then:
\begin{codeblock}
class node \{
public:
    // constructors, destructors, delegators,
    // etc ...
private:
    decoration *deco;
    multiset< pair< node, bool > > children;
\};
\end{codeblock}

Note that a {\tt node} can be either the entire tree, or a subtree or
a single (atomic) leaf.  The layout of class {\tt decoration} holds a
pair of indices and in the case of QED a GiNaC expression (class {\tt
  ex}) for the gauge.  An object of class {\tt ex} is entirely
sufficient, since it may contain either a numerical value (integer or
fractional) or a symbol (like \(\xi\)) (or even more complex
expressions, if need should arise).

The knowledge how to manipulate the indices \(i\) and \(j\), depending
on the type of decoration and on the state of the edges is built into
the classes of type {\tt decoration}.  They are automatically adjusted
when the tree is constructed.  Furthermore, the state of the edges is
also irrelevant for the user of the program since we need all possible
combinations of cuts.  If there are \(n\) vertices inside a tree, then
there are \(n-1\) edges and we have to construct all possible
\(2^{n-1}\) combinations.

Once all the trees have been created we call a method called {\tt
  evaluate} in each of them that traverses the tree in a top-down
fashion: In each node, the expression in terms of
\(F_{a+i\eps,b+j\eps}\) is dispatched and expanded as a series in the
regularization parameter \(\eps\) before {\tt evaluate} is called on
the multiset of children.  The resulting \(2^{n-1}\) Laurent series
are then added together and the coefficients are expanded. This
procedure yields the antipode.

Here is an example how the programs are used in practice:
\begin{codeblock}
\$ ./qed1 "(((Sigma[0])(Sigma[xi])Gamma[-1])Gamma[xi])"
After decoration the tree has these indices:
  (Gamma[3,0][xi](Gamma[2,0][-1](Sigma[0,0][0])(Sigma[0,0][xi])))
----+----#----+----#----+----#----+----#----+----#----+----#
The antipode of this tree is:

   (-1/4-1/4*xi^2-1/2*xi)*x^(-3)
  +(7/24+1/24*xi^2+1/3*xi)*x^(-2)
  +(-5/16-13/48*xi^2-7/12*xi)*x^(-1)
\end{codeblock}

The graph to be computed is passed in tree form as a string on the
command line, together with the gauge parameter.  Note how the
indices \(i\) and \(j\) are then set up automatically in line 3.  The
next line is a simple progress bar, useful when computations become
longer.  The result is then printed as a power series in the
regularization parameter, called {\tt x} instead of \(\eps\) in order
to please the computer.

Computationally, the results are always quite small (c.f.\
section~\ref{sec:results}) but there is a huge swell of intermediate
expressions. After all, the Laurent series arise from expanding
products and sums of lots of \(\Gamma\)-functions and so inevitably
carry Euler-Mascheroni constants. But these have to vanish by
cancellations, a fact that is conveniently used as a consistency check
for the result. In practice, antipodes of nine-loop graphs in massless
QED are still computationally feasible but may require several hours
and hundreds of megabytes memory. This emphasizes the drastic loss of
efficiency in comparison with~\cite{BrKr,BrKrff}, which is compensated
by a gain in flexibility to handle different decorations at any step
in the recursion.

The programs developed herein are written in plain ISO
C++~\cite{C++Standard} and will run on any system where the GiNaC
library has been ported to.  Note that porting to a new platform will
require porting CLN~\cite{CLN} to that platform first, since GiNaC
depends on CLN for its arbitrary precision arithmetic.

\section{Computational Results}\label{sec:results}


In a first step, to examine the appearance of $\zeta$-functions,
we will now calculate the antipode of different graphs given in
figure~\ref{fig:graphs_trees_yukawa} and
figure~\ref{fig:graphs_trees_qed} that represent the two
topologies of figure~\ref{fig:ladder_and_cheese}. In Yukawa theory
the graphs $\Gamma^1_{[1,1,1]}$ and $\Gamma^2_{[1,1,1]}$ of
figure~\ref{fig:graphs_trees_yukawa} are representations of the
ladder topology~\ref{fig:ladder_and_cheese}.b, the graphs
$\Gamma^1_{[2,0,1]}$ and $\Gamma^2_{[2,0,1]}$ of the swiss cheese
topology~\ref{fig:ladder_and_cheese}.a. The same holds for QED
with $[1,1,1]$ replaced by $[1,1,2]$ and $[2,0,1]$ by $[2,0,2]$.

The graphs for the Yukawa theory, together with their expressions
in characterizing functions are given in
figure~\ref{fig:graphs_trees_yukawa}. The figure also contains the
decorated rooted trees we get from them to calculate the antipode.
One can see that except from the decorations they all belong to
the same rooted tree, and hence have the same structure as far as
their short distance singularities go, as advertised in the
introduction.

\begin{figure}[t]
\SetScale{1}
\begin{center}
\begin{tabular}{llll}
\hspace{-1cm}
$\Gamma^1_{[1,1,1]}$
&
$\equiv$
&
\quad
\SetScale{0.5}
\begin{picture}(150,10)(0,0)
\Line(160,50)(40,0)
\Line(40,0)(160,-50)
\DashLine(0,0)(40,0){4}
\DashLine(143,43)(143,-43){4}
\DashLine(115,31)(115,-31){4}
\DashArrowArc(73,-14)(11,-22,158){4}
\BCirc(115,0){9.5}
\end{picture}
&
\hspace{-1cm}$\equiv \quad (-ig)^9\Gamma^1_{3,0}\Gamma^1_{1,1}\Sigma_{0,0}\Pi_{0,0}[q^2]^{-4\eps}$\\
&\\&\\&\\\end{tabular}\end{center}
\hspace{1cm}
\begin{center}\begin{tabular}{cccc}
\hspace{-1cm}
$\Gamma^2_{[1,1,1]}$
&
$\equiv$
&
\quad
\SetScale{0.5}
\begin{picture}(150,10)(0,0)
\DashLine(160,50)(145,44){4}
\Line(145,44)(115,31)
\DashLine(115,31)(40,0){4}
\DashLine(115,-31)(145,-44){4}
\Line(40,0)(115,-31)
\Line(145,-44)(160,-50)
\Line(0,0)(40,0)
\Line(145,44)(145,-44)
\Line(115,31)(115,-31)
\DashArrowArc(73,-14)(11,-22,158){4}
\DashArrowArc(115,0)(11,270,90){4}
\end{picture}
&
\hspace{-1cm}$\equiv \quad (-ig)^9\Gamma^2_{3,0}\Gamma^2_{1,1}\Sigma_{0,0}\Sigma_{0,0}[q^2]^{-4\eps}$\\
&\\&\\&\\\end{tabular}\end{center}
\begin{center}
\begin{tabular}{cccc}
\hspace{-1cm}
$\Gamma^1_{[2,0,1]}$
&
$\equiv$
&
\quad
\SetScale{0.5}
\begin{picture}(150,30)(0,0)
\Line(160,50)(40,0)
\Line(40,0)(160,-50)
\DashLine(0,0)(40,0){4}
\DashLine(145,44)(145,-44){4}
\DashLine(124,35)(124,-35){4}
\DashArrowArc(63,-10)(11,-22,158){4}
\DashArrowArc(95,-22)(11,-22,158){4}
\end{picture}
&
\hspace{-1cm}$\equiv \quad (-ig)^9\Gamma^1_{3,0}\Gamma^1_{2,0}\Sigma_{0,0}\Sigma_{0,0}[q^2]^{-4\eps}$
\\&\\&\\&\\
\end{tabular}\end{center}
\hspace{1cm}
\begin{center}
\begin{tabular}{cccc}
\hspace{-1cm}
$\Gamma^2_{[2,0,1]}$
&
$\equiv$
&
\quad
\SetScale{0.5}
\begin{picture}(150,10)(0,0)
\DashLine(160,50)(145,44){4}
\Line(145,44)(124,35)
\DashLine(124,35)(40,0){4}
\DashLine(124,-35)(145,-44){4}
\Line(40,0)(124,-35)
\Line(145,-44)(160,-50)
\Line(0,0)(40,0)
\Line(145,44)(145,-44)
\Line(124,35)(124,-35)
\DashArrowArc(63,-10)(11,-22,158){4}
\DashArrowArc(95,-22)(11,-22,158){4}
\end{picture}
&
\hspace{-1cm}$\equiv \quad (-ig)^9\Gamma^2_{3,0}\Gamma^2_{2,0}\Sigma_{0,0}\Sigma_{0,0}[q^2]^{-4\eps}$
\\&\\&\\&\\
\end{tabular}\end{center}
\SetScale{1}
\begin{center}
\begin{tabular}{cc}
\SetScale{1}
\begin{picture}(50,50)(0,0)
\Vertex(10,-20){5}
\Vertex(40,-20){5}
\Vertex(25,0){5}
\Vertex(25,22){5}
\Line(10,-20)(25,0)
\Line(25,0)(40,-20)
\Line(25,0)(25,22)
\SetScale{0.2}
\Line(-110,-120)(-10,-120)
\DashArrowArc(-60,-120)(40,0,180){4}
\SetScale{1}
\SetScale{0.2}
\Line(250,-120)(350,-120)
\DashArrowArc(300,-120)(40,0,180){4}
\Line(365,155)(255,115)
\Line(255,115)(365,75)
\DashLine(220,115)(255,115){4}
\DashLine(335,144)(335,86){4}
\Line(365,45)(255,5)
\Line(255,5)(365,-35)
\DashLine(220,5)(255,5){4}
\DashLine(335,33)(335,-23){4}
\SetScale{1}
\end{picture}
\hspace{4cm}
&
\begin{picture}(50,50)(0,0)
\Vertex(10,-20){5}
\Vertex(40,-20){5}
\Vertex(25,0){5}
\Vertex(25,22){5}
\Line(10,-20)(25,0)
\Line(25,0)(40,-20)
\Line(25,0)(25,22)
\SetScale{0.2}
\Line(-110,-120)(-10,-120)
\DashArrowArc(-60,-120)(40,0,180){4}
\SetScale{1}
\SetScale{0.2}
\BCirc(310,-100){29}
\DashLine(255,-100)(281,-100){4}
\DashLine(339,-100)(365,-100){4}
\Line(365,155)(255,115)
\Line(255,115)(365,75)
\DashLine(220,115)(255,115){4}
\DashLine(335,144)(335,86){4}
\Line(365,45)(255,5)
\Line(255,5)(365,-35)
\DashLine(220,5)(255,5){4}
\DashLine(335,33)(335,-23){4}
\SetScale{1}
\end{picture}
\end{tabular}
\end{center}
\vspace{1cm}
\caption{\small Graphs in Yukawa theory and their decorated rooted trees.}
\label{fig:graphs_trees_yukawa}
\end{figure}

\begin{figure}[t]
\begin{tabular}{ccc}
$\Gamma^1_{[1,1,2]}\:\equiv$
&
\quad
\SetScale{0.5}
\begin{picture}(150,50)(0,0)
\Line(172,60)(40,0)
\Line(40,0)(172,-60)
\Photon(0,0)(40,0){4}{5.5}
\Photon(162,56)(162,-56){4}{11.5}
\Photon(138,45)(138,-45){4}{9.5}
\Photon(112,32)(112,-32){4}{7.5}
\PhotonArc(73,-14)(11,-24,156){2}{5.5}
\BCirc(112,0){9.5}
\end{picture}
\SetScale{1}
&
$\equiv\quad (-ie)^{11}M^1_{4,0}M^1_{3,0}M^1_{1,1}\tilde{\Sigma}_{0,0}\Pi_{0,0}[q^2]^{-5\eps}$
\\&&\\&&\\
$\Gamma^2_{[1,1,2]}\:\equiv$
&
\quad
\SetScale{0.5}
\begin{picture}(150,50)(0,0)
\Photon(40,0)(105,30){4}{7}
\Line(105,30)(141,47)
\Photon(141,47)(162,56){4}{2.5}
\Line(162,56)(172,60)
\Line(40,0)(105,-30)
\Photon(105,-30)(141,-47){4}{4.5}
\Line(141,-47)(162,-56)
\Photon(162,-56)(172,-60){4}{1.5}
\Line(0,0)(40,0)
\Line(162,56)(162,-56)
\Line(141,47)(141,-47)
\Line(105,30)(105,-30)
\PhotonArc(73,-14)(11,-24,156){2}{5.5}
\PhotonArc(105,0)(11,270,90){2}{5.5}
\end{picture}
\SetScale{1}
&
$\equiv\quad (-ie)^{11}M^2_{4,0}M^2_{3,0}M^2_{1,1}\tilde{\Sigma}_{0,0}\tilde{\Sigma}_{0,0}[q^2]^{-5\eps}$
\\&&\\&&\\&&\\
$\Gamma^1_{[2,0,2]}\:\equiv$
&
\quad
\SetScale{0.5}
\begin{picture}(150,50)(0,0)
\Line(172,60)(40,0)
\Line(40,0)(172,-60)
\Photon(0,0)(40,0){4}{5.5}
\Photon(162,56)(162,-56){4}{11.5}
\Photon(141,47)(141,-47){4}{9.5}
\Photon(120,37)(120,-37){4}{7.5}
\PhotonArc(63,-10)(11,-24,156){2}{5.5}
\PhotonArc(93,-24)(11,-24,156){2}{5.5}
\end{picture}
\SetScale{1}
&
$\equiv\quad (-ie)^{11}
M^1_{4,0}M^1_{3,0}M^1_{2,0}\tilde{\Sigma}_{0,0}\tilde{\Sigma}_{0,0}[q^2]^{-5\eps}
$
\\&&\\&&\\
$\Gamma^2_{[2,0,2]}\:
\equiv$
&
\quad
\SetScale{0.5}
\begin{picture}(150,50)(0,0)
\Photon(40,0)(120,37){4}{7}
\Line(120,37)(141,47)
\Photon(141,47)(162,56){4}{2.5}
\Line(162,56)(172,60)
\Line(40,0)(120,-37)
\Photon(120,-37)(141,-47){4}{2.5}
\Line(141,-47)(162,-56)
\Photon(162,-56)(172,-60){4}{1.5}
\Line(0,0)(40,0)
\Line(162,56)(162,-56)
\Line(141,47)(141,-47)
\Line(120,37)(120,-37)
\PhotonArc(63,-10)(11,-24,156){2}{5.5}
\PhotonArc(93,-24)(11,-24,156){2}{5.5}
\end{picture}
\SetScale{1}
&
$\equiv\quad (-ie)^{11}
M^2_{4,0}M^2_{3,0}M^2_{2,0}\tilde{\Sigma}_{0,0}\tilde{\Sigma}_{0,0}[q^2]^{-5\eps}
$
\\&&\\&&\\
\end{tabular}\\
\\
\SetScale{1}
\begin{center}
\begin{tabular}{cc}
\SetScale{1}
\begin{picture}(50,50)(0,0)
\Vertex(10,-20){5}
\Vertex(40,-20){5}
\Vertex(25,0){5}
\Vertex(25,22){5}
\Vertex(25,44){5}
\Line(10,-20)(25,0)
\Line(25,0)(40,-20)
\Line(25,0)(25,22)
\Line(25,22)(25,44)
\SetScale{0.2}
\Line(-110,-120)(-10,-120)
\PhotonArc(-60,-120)(40,0,180){4}{9.5}
\SetScale{1}
\SetScale{0.2}
\Line(250,-120)(350,-120)
\PhotonArc(300,-120)(40,0,180){4}{9.5}
\Line(365,265)(255,225)
\Line(255,225)(365,185)
\Photon(220,225)(255,225){4}{6.5}
\Photon(335,254)(335,196){4}{6.5}
\Line(365,155)(255,115)
\Line(255,115)(365,75)
\Photon(220,115)(255,115){4}{6.5}
\Photon(335,144)(335,86){4}{6.5}
\Line(365,45)(255,5)
\Line(255,5)(365,-35)
\Photon(220,5)(255,5){4}{6.5}
\Photon(335,33)(335,-23){4}{5.5}
\SetScale{1}
\SetScale{1}
\end{picture}
\hspace{4cm}
&
\begin{picture}(50,50)(0,0)
\Vertex(10,-20){5}
\Vertex(40,-20){5}
\Vertex(25,0){5}
\Vertex(25,22){5}
\Vertex(25,44){5}
\Line(10,-20)(25,0)
\Line(25,0)(40,-20)
\Line(25,0)(25,22)
\Line(25,22)(25,44)
\SetScale{0.2}
\Line(-110,-120)(-10,-120)
\PhotonArc(-60,-120)(40,0,180){4}{9.5}
\SetScale{1}
\SetScale{0.2}
\Photon(250,-100)(291,-100){4}{6.5}
\Photon(349,-100)(390,-100){4}{6.5}
\BCirc(320,-100){29}
\Line(365,265)(255,225)
\Line(255,225)(365,185)
\Photon(220,225)(255,225){4}{6.5}
\Photon(335,254)(335,196){4}{6.5}
\Line(365,155)(255,115)
\Line(255,115)(365,75)
\Photon(220,115)(255,115){4}{6.5}
\Photon(335,144)(335,86){4}{6.5}
\Line(365,45)(255,5)
\Line(255,5)(365,-35)
\Photon(220,5)(255,5){4}{6.5}
\Photon(335,33)(335,-23){4}{5.5}
\SetScale{1}
\SetScale{1}
\end{picture}
\end{tabular}
\end{center}
\vspace{1cm}
\caption{\small Graphs in QED and their decorated rooted trees.}
\label{fig:graphs_trees_qed}
\end{figure}

In QED, the considered graphs in figure~\ref{fig:graphs_trees_qed}
possess one more vertex correction compared to the graphs in yukawa
theory. This additional photon line is necessary because the short
distance singularity structure is actually determined by the bidegree,
ie.\ the number of subdivergent graphs. It so happens in QED that the
one-loop fermion self-energy and one-loop vertex corrections (with divergent subgraphs) are
overall finite if the internal photon is transversal. In our case, we thus
find that the insertion of a one-loop vacuum polarization into the
internal photon line in a vertex results in a convergent vertex
correction. We have to plug the whole function into one more vertex
correction to get to the next level in the bidegree. And indeed, we
need an additional vertex correction, compared to Yukawa theory, to
obtain the $\zeta(3)$ in the pole-terms: gauge symmetry delays the
appearance of transcendentals~\cite{pisa}. Also let us mention that we
can easily compare graphs which have an internal vacuum polarization
with graphs which have an internal fermion self-energy by using the
before-mentioned elimination of the trace in the vacuum polarization,
using
\begin{eqnarray*}
\Pi_{0,0}=\textrm{tr}(\Id)\,\frac{1}{2}\,F_{1,1}=\textrm{tr}(\Id)\,\Sigma_{0,0}.
\end{eqnarray*}
Furthermore, $\Gamma^2_{[1,1,2]}$ and $\Gamma^2_{[2,0,2]}$ in QED do
not have this extra shift between loop number and bidegree by
themselves, as there the desire to maintain the same topology never
forces us to use a vacuum polarization as a subgraph: all one-loop
subdivergences are fermion self-energies and vertex corrections.
However, we use the same gauges in these graphs as in the graphs
$\Gamma^1_{[1,1,2]}$ and $\Gamma^1_{[2,0,2]}$ to compare the results
directly (see below), and therefore obtain the same difference between
loop-number and bidegree.

Finally, in our results, we are only interested in residues of
counterterms, poles of first order. All interesting relations
between transcendental degree and topology will appear there. The
scattering type formula~\cite{RHII} will make sure that parts of
these relations will then resurface in the poles of higher order
in the counterterm, but they contain no new information. So in the
following we solely exhibit the residues of the counterterms for
our selected class of graphs. We denote by ${\rm res}(\Gamma)$
this coefficient of the pole of first order in the MS counterterm
in dimensional regularization of a graph $\Gamma$.
Assorted results for these residues  are:
\pagebreak

\noindent Yukawa theory:\\

\noindent\begin{tabular*}{\textwidth}{@{}p{.12\textwidth}p{.02\textwidth}p{.84\textwidth}@{}}
\({\rm res} (\Gamma^1_{[1,1,1]})\)&\(=\)&\(\displaystyle \frac{5}{48}\) \\[1em]
\({\rm res} (\Gamma^2_{[1,1,1]})\)&\(=\)&\(\displaystyle \frac{1}{24}\) \\[1em]
\({\rm res} (\Gamma^1_{[2,0,1]})\)&\(=\)&\(\displaystyle \left(\frac{5}{48}-\frac{1}{8}\zeta(3)\right)\) \\[1em]
\({\rm res}(\Gamma^2_{[2,0,1]})\)&\(=\)&\(\displaystyle \left(\frac{1}{12}-\frac{1}{8}\zeta(3)\right)\)
\\[2.5em]
\end{tabular*}

\noindent QED:\\

\noindent\begin{tabular*}{\textwidth}{@{}p{.12\textwidth}p{.02\textwidth}p{.84\textwidth}@{}}
\({\rm res} (\Gamma^1_{[1,1,2]})\)&\(=\)&
\(\displaystyle -\left(\frac{1}{480}(301+143\xi-170\xi^2)\right)\)
\\
\({\rm res} (\Gamma^2_{[1,1,2]})\)&\(=\)&
\(\displaystyle -\left(\frac{1}{960}(584+659\xi+59\xi^2)
\right)\)
\\
\({\rm res} (\Gamma^1_{[2,0,2]})\)&\(=\)&
\(\displaystyle -
\left(
\frac{1}{480}(521+309\xi-236\xi^2)
+\frac{3}{10}\zeta(3)(1+\xi)^2  \right)\)
 \\
\({\rm res}
(\Gamma^2_{[2,0,2]})\)&\(=\)&
\(\displaystyle -
\left(\frac{1}{960}(832+879\xi+31\xi^2)+\frac{3}{10}\zeta(3)(1+\xi)^2\right)\)
\\[1.5em]
\end{tabular*}

We see, that in Yukawa theory the graphs $\Gamma^1_{[2,0,1]}$ and
$\Gamma^2_{[2,0,1]}$ with the swiss cheese topology have a residue
involving $\zeta(3)$, while $\Gamma^1_{[1,1,1]}$ and
$\Gamma^2_{[1,1,1]}$ realizing the ladder topology just have a
rational residue, as expected. A similar result holds in QED: the
graphs $\Gamma^1_{[2,0,2]}$ and $\Gamma^2_{[2,0,2]}$ with the
swiss cheese topology again have a residue involving $\zeta(3)$,
while $\Gamma^1_{[1,1,2]}$ and $\Gamma^2_{[1,1,2]}$ realizing the
ladder topology just have a rational residue.
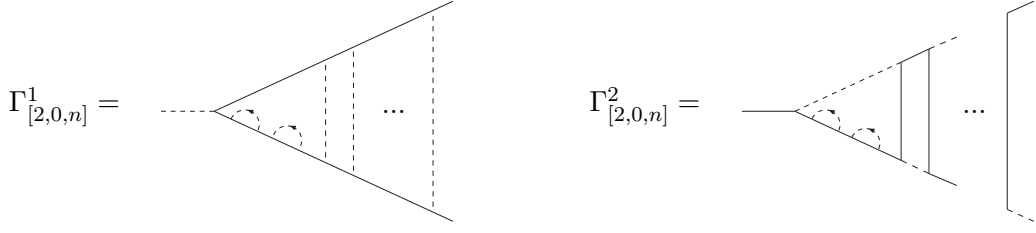
\begin{figure}[t]
\begin{center}
\begin{tabular}{cccc}
$\Gamma^1_{[2,0,n]}=$
&
\SetScale{0.5}
\begin{picture}(150,20)(0,0)
\Line(220,83)(40,0)
\Line(40,0)(220,-83)
\DashLine(0,0)(40,0){4}
\DashLine(145,45)(145,-45){4}
\DashLine(124,35)(124,-35){4}
\DashLine(205,72)(205,-72){4}
\DashArrowArc(63,-10)(11,-22,158){4}
\DashArrowArc(95,-22)(11,-22,158){4}
\Text(88,0)[]{...}
\end{picture}
&
$\Gamma^2_{[2,0,n]}=$
&
\SetScale{0.5}
\begin{picture}(150,20)(0,0)
\Line(200,74)(220,83)
\DashLine(200,-74)(220,-83){4}
\Line(200,-74)(200,74)
\DashLine(40,0)(120,37){4}
\Line(120,37)(141,47)
\DashLine(141,47)(162,56){4}
\Line(40,0)(120,-37)
\DashLine(120,-37)(141,-47){4}
\Line(141,-47)(162,-56)
\Line(0,0)(40,0)
\Line(141,47)(141,-47)
\Line(120,37)(120,-37)
\DashArrowArc(63,-10)(11,-24,156){4}
\DashArrowArc(93,-24)(11,-24,156){4}
\Text(88,0)[]{...}
\end{picture}
\SetScale{1}
\end{tabular}
\vspace{1.7cm}
\end{center}
\caption{\small Iterated vertex corrections.}
\label{fig:iterated_vertex}
\end{figure}
Those residues are in general a linear combination of terms of varying
transcendental weight. This weight vanishes for a rational number, and
in accordance with standard practice in the study of multiple zeta
values the transcendental weight $w$ of a monomial $\prod_i
\zeta(j_i)$ is $$w(\prod_i \zeta(j_i))=\sum_i j_i.$$
Then, the above
results confirm that the coefficient of the highest weight
transcendental in the transition from ${\rm res}(\Gamma^1_{[2,0,1]})$
to ${\rm res}(\Gamma^2_{[2,0,1]})$ (${\rm res}(\Gamma^1_{[2,0,2]})$ to
${\rm res}(\Gamma^2_{[2,0,2]})$ for QED) is invariant. This is the
partial symmetry announced in the introduction. It holds for any gauge
in the QED case and holds in general, if we increase the loop number
as for example in figure (\ref{fig:iterated_vertex}).

\noindent In Yukawa theory we get:\\

\noindent\begin{tabular*}{\textwidth}{@{}p{.12\textwidth}p{.02\textwidth}p{.84\textwidth}@{}}
\({\rm res }(\Gamma^1_{[2,0,1]})\) &\(=\)&
\(\displaystyle\left(\frac{5}{48}-\frac{1}{8}\zeta(3)\right)\)
\\[1em]
\({\rm res }(\Gamma^2_{[2,0,1]})\)&\(=\)&
\(\displaystyle\left(\frac{1}{12}-\frac{1}{8}\zeta(3)\right)\)
\\[1em]
\end{tabular*}

\noindent\begin{tabular*}{\textwidth}{@{}p{.12\textwidth}p{.02\textwidth}p{.84\textwidth}@{}}
\({\rm res }(\Gamma^1_{[2,0,2]})\)&\(=\)&
\(\displaystyle\left(\frac{1}{20}-\frac{9}{40}\zeta(3)-\frac{3}{80}\zeta(4)\right)\)
\\[1em]
\({\rm res }(\Gamma^2_{[2,0,2]})\)&\(=\)&
\(\displaystyle\left(\frac{1}{240}-\frac{1}{20}\zeta(3)-\frac{3}{80}\zeta(4)\right)\)
\\[1em]
\end{tabular*}

\noindent\begin{tabular*}{\textwidth}{@{}p{.12\textwidth}p{.02\textwidth}p{.84\textwidth}@{}}
\({\rm res }(\Gamma^1_{[2,0,3]})\)&\(=\)&
\(\displaystyle
-
\left(
\frac{23}{90}
+\frac{9}{20}\zeta(3)
+\frac{7}{80}\zeta(4)
+\frac{7}{240}\zeta(5)
\right)\)
\\[1em]
\({\rm res }(\Gamma^2_{[2,0,3]})\) &\(=\)&
\(\displaystyle
-
\left(
-\frac{1}{240}
+\frac{1}{6}\zeta(3)
+\frac{1}{80}\zeta(4)
+\frac{7}{240}\zeta(5)
\right)\)
\\[1em]
\end{tabular*}

\noindent\begin{tabular*}{\textwidth}{@{}p{.12\textwidth}p{.02\textwidth}p{.84\textwidth}@{}}
\({\rm res }(\Gamma^1_{[2,0,4]})\)&\(=\)&
\(\displaystyle-
\left(
\frac{919}{630}
+\frac{71}{70}\zeta(3)
+\frac{111}{560}\zeta(4)
+\frac{1}{12}\zeta(5)
+\frac{1}{560}\zeta(3)^2
+\frac{1}{112}\zeta(6)
\right)\)
\\[1em]
\({\rm res }(\Gamma^2_{[2,0,4]})\)&\(=\)&
\(\displaystyle-
\left(
\frac{65}{224}
+\frac{11}{140}\zeta(3)
+\frac{1}{16}\zeta(4)
+\frac{1}{120}\zeta(5)
+\frac{1}{560}\zeta(3)^2
+\frac{1}{112}\zeta(6)
\right)\)
\\[1em]
\end{tabular*}

\noindent
\begin{tabular*}{\textwidth}{@{}p{.12\textwidth}p{.02\textwidth}p{.84\textwidth}@{}}
\({\rm res }(\Gamma^1_{[2,0,5]})\)&\(=\)&
\(\displaystyle-
\left(
\frac{6481}{1120}
+\frac{33613}{13440}\zeta(3)
+\frac{2133}{4480}\zeta(4)
+\frac{101}{480}\zeta(5)
+\frac{27}{4480}\zeta(3)^2
\right.\)
\\[.8em]
&&
\(\displaystyle\left.\quad\quad\quad
+\frac{27}{896}\zeta(6)
+\frac{3}{4480}\zeta(3)\zeta(4)
+\frac{7}{1920}\zeta(7)
\right)\)
\\[1em]
\({\rm res }(\Gamma^2_{[2,0,5]})\)&\(=\)&
\(\displaystyle-
\left(
\frac{863}{3360}
+\frac{61}{160}\zeta(3)
+\frac{27}{1120}\zeta(4)
+\frac{7}{120}\zeta(5)
+\frac{1}{2240}\zeta(3)^2
\right.\)
\\[.8em]
&&
\(\displaystyle\left.\quad\quad\quad
+\frac{1}{448}\zeta(6)
+\frac{3}{4480}\zeta(3)\zeta(4)
+\frac{7}{1920}\zeta(7)
\right)\)
\end{tabular*}

\noindent Similarly, for QED we find:\\

\noindent\begin{tabular*}{\textwidth}{@{}p{.12\textwidth}p{.02\textwidth}p{.84\textwidth}@{}}
\({\rm res }(\Gamma^1_{[2,0,2]})\) &\(=\)&
\(\displaystyle-\frac{1}{480}(521+309\xi-236\xi^2)-\frac{3}{10}(1+\xi)^2\zeta(3)\)
\\[1em]
\({\rm res }(\Gamma^2_{[2,0,2]})\)&\(=\)&
\(\displaystyle-\frac{1}{960}(832+879\xi+31\xi^2)-\frac{3}{10}(1+\xi)^2\zeta(3)\)
\\[1em]
\end{tabular*}

\noindent\begin{tabular*}{\textwidth}{@{}p{.12\textwidth}p{.02\textwidth}p{.84\textwidth}@{}}
\({\rm res }(\Gamma^1_{[2,0,3]})\)&\(=\)&
\(\displaystyle-\frac{1}{5760}(13377+16773\xi+12091\xi^2+9463\xi^3)+\frac{1}{30}(1+\xi)^2(-38+7\xi)\zeta(3)\)
\\[.8em]
&&
\(\displaystyle-\frac{3}{40}(1+\xi)^3\zeta(4)\)
\\[1em]
\({\rm res }(\Gamma^2_{[2,0,3]})\)&\(=\)&
\(\displaystyle-\frac{1}{5760}(13327+32578\xi+26038\xi^2+6811\xi^3)
-\frac{1}{60}(1+\xi)^2(31+4\xi)\zeta(3)\)
\\[.8em]
&&
\(\displaystyle-\frac{3}{40}(1+\xi)^3\zeta(4)\)
\\[1em]
\end{tabular*}

\noindent\begin{tabular*}{\textwidth}{@{}p{.12\textwidth}p{.02\textwidth}p{.84\textwidth}@{}}
\({\rm res }(\Gamma^1_{[2,0,4]})\)&\(=\)&
\(\displaystyle-\frac{1}{40320}(427681+1048812\xi+889375\xi^2+130774\xi^3-145630\xi^4)\)
\\[.8em]
&&
\(\displaystyle\quad
-\frac{1}{840}(1+\xi)^2(2551+425\xi+736\xi^2)\zeta(3)
+\frac{1}{280}(1+\xi)^3(-121+23\xi)\zeta(4)\)
\\[.8em]
&&
\(\displaystyle\quad
-\frac{1}{20}(1+\xi)^4\zeta(5)\)
\\[1em]
\({\rm res }(\Gamma^2_{[2,0,4]})\) &\(=\)&
\(\displaystyle-\frac{1}{161280}(1357764+4016396\xi+4028301\xi^2+1438370\xi^3+68749\xi^4)\)
\\[.8em]
&&
\(\displaystyle\quad
-\frac{1}{1680}(1+\xi)^2(1448+2365\xi+836\xi^2)\zeta(3)
-\frac{1}{140}(1+\xi)^3(29+2\xi)\zeta(4)\)
\\[.8em]
&&
\(\displaystyle\quad
-\frac{1}{20}(1+\xi)^4\zeta(5)\)
\\[1em]
\end{tabular*}

\noindent\begin{tabular*}{\textwidth}{@{}p{.12\textwidth}p{.02\textwidth}p{.84\textwidth}@{}}
\({\rm res }(\Gamma^1_{[2,0,5]})\)&\(=\)&
\(\displaystyle\frac{1}{430295040}(-19150607852-55087254529\xi-55376777218\xi^2-24306202368\xi^3\)
\\[.8em]
&&
\(\displaystyle\hspace{3cm}-10014576786\xi^4-5440770359\xi^5)\)
\\[.8em]
&&
\(\displaystyle+\frac{1}{4480}(1+\xi)^2(-39057-40975\xi-19389\xi^2+6061\xi^3)\zeta(3)\)
\\[.8em]
&&
\(\displaystyle-\frac{3}{4480}(1+\xi)^3(2031+11\xi+446\xi^2)\zeta(4)
+\frac{1}{960}(1+\xi)^4(-347+67\xi)\zeta(5)\)
\\[.8em]
&&
\(\displaystyle-\frac{3}{1120}(1+\xi)^5\zeta(3)^2-\frac{3}{224}(1+\xi)^5\zeta(6)\)
\\[1em]
\({\rm res }(\Gamma^2_{[2,0,5]})\)&\(=\)&
\(\displaystyle-\frac{1}{1290240}(29658556+111493999\xi+161539373\xi^2+112812993\xi^3\)
\\[.8em]
&&
\(\displaystyle\hspace{3cm}+38688379\xi^4+5579364\xi^5)\)
\\[.8em]
&&
\(\displaystyle-\frac{1}{26880}(1+\xi)^2(80770+155550\xi +79521\xi^2+3472\xi^3)\zeta(3)\)
\\[.8em]
&&
\(\displaystyle-\frac{1}{8960}(1 + \xi)^3(3254+3919\xi+1340\xi^2)\zeta(4)\)
\\[.8em]
&&
\(\displaystyle-\frac{1}{960}(1+\xi)^4(179+8\xi)\zeta(5)\)
\\[.8em]
&&
\(\displaystyle-\frac{3}{1120}(1+\xi)^5\zeta(3)^2
-\frac{3}{224}(1+\xi)^5\zeta(6)\)
\\[1em]
\end{tabular*}

The reader will note that in QED our residues are polynomials in the gauge
parameter of a degree reduced by two steps from what one might expect,
to enable comparison between configurations with insertions of
self-energies into photon or fermion lines. The corresponding fermion
self-energy was for that purpose evaluated in the Feynman gauge, and
in the Landau gauge for the affected photon propagator. If one only
compares cases with self-energy insertions at fermionic lines, one can
abandon these restrictions and we did confirm that the reported
invariance holds as expected with coefficients which are polynomials
in the gauge parameter of degree equal to the number of photon lines.

\section{Discussion and Proof}\label{sec:discussion}
Actually, results of the form reported in the previous sections
can be derived from the analytic structure of the functions
$F_{a,b}$ and some basic field theoretic arguments.

Let us reconsider the situation. The simplicity of Feynman graphs
considered here manifests itself computationally by the fact that
they factorize in a unique manner. Each divergent subgraph
$\gamma$ depends only on a single external momentum $q$ say (that
is the reason why we only consider vertex subgraphs at zmt), such
that its evaluation in dimensional regularization gives a result
of the form
$$ \phi(\gamma)(\eps)=[q^2]^{-n(\gamma)\eps}\sum_i F_i(\eps)c_i(q).$$
Here, $n(\gamma)$ is the number of loops in $\gamma$, and
$F_i(\eps)$ are $q$-independent form factors, and the dimensionless
$c_i(q)$ are\\
$c_1(q)=\gamma_\mu,c_2(q)=q_\mu q\hspace{-1.8mm}{/}/q^2$
for the QED vertex (the only case here in which the sum has more
than one term),\\ $c_1(q)=1$ for the vertex correction in Yukawa
theory,\\  $c_1(q)= q\hspace{-1.8mm}{/}$ for any fermion self-energy,\\
$c_1(q)=g_{\mu\nu}-q_\mu q_\nu/q^2$ for the photon,\\ $c_1(q)=1$
for the scalar boson.

Insertions of such graphs $\gamma$ in another graph $\Gamma$ only
raises powers of the scalar part of some propagator of $\Gamma$:
$$
\frac{1}{q^2}\to \frac{1}{[q^2]^{1+n(\gamma)\eps}}.$$  We can keep
track of this by notating these loop numbers $n(\gamma)$ in the
entries $M(\Gamma)_{ij}$ for the corresponding propagator in the adjacency matrix
$M(\Gamma)$.

Let us now assume that $\Gamma$ is some primitive vertex
correction, ie.\ free of divergent subgraphs, and let us write as
before $\Gamma^1$ and $\Gamma^2$ for two distinct choices of zmt.

Consider a bunch of 1PI graphs $X=\prod_{i=1}^k \gamma_i$ each
dependent on a sole external momentum as described above. Let
$k={\rm bid}(X)$ be the bidegree of $X$, so that $X$ has a highest
pole in $\eps$ of degree $k$ with coefficient $c^X_k$. Let now
$G_X$ be chosen gluing data such that
$\Gamma_X:=\Gamma\star_{G_X}X$ is obtained from inserting $X$ at
specified vertices and propagators into $\Gamma$, with ${\rm
bid}(\Gamma)=1$ without loss of generality. (Any 1PI graph can be
written in the form $\Gamma\star_{G_X}X$ for appropriate such
$\Gamma,X$~\cite{Hopfalgebra,Overlapping,RHII}, in generalization
of the closed Hochschild one cocycle $B_+$ of undecorated rooted
trees). Further,  each $X$ allows for an expansion
$$ \phi(X)=\frac{c^X_k}{\eps^k}(1+T(X)(\eps))$$
and similarly, let $\phi(\Gamma)= \frac{{\rm
res}(\Gamma)}{\eps}(1+T(\Gamma)(\eps))$. Now assume that the
Taylor series \linebreak
$[1+T(X)(\eps)][1+T(\Gamma)(\eps)]=1+\sum_{j=1}^\infty c_j\eps^j$
is such that the transcendental weight $w(c_j)$ of $c_j$ increases
with $j$:
$$ w(c_j)<w(c_{j+1}),$$
$\forall j\geq k-1$. Here, we define the transcendental weight of an
expression which is a sum of terms as the highest transcendental
weight appearing in its terms.\footnote{The question as to how define the
transcendental weight in a context which exceeds the Riemann
$\zeta$-functions or MZVs~\cite{pisa} we do not have to answer
here. Also, the attentive reader might have noticed that we set
the transcendental weight of the gauge parameter to be zero for
the QED results, treating it as an independent variable.}
\begin{prop}
The counterterm is the same for $\Gamma_X^1$ and $\Gamma_X^2$,
and hence their residues are equal.
\end{prop}
Proof: Elementary, as $\Gamma^1-\Gamma^2$ is UV convergent,
and hence $\Gamma^1_X$ and $\Gamma^2_X$ generate the same counterterm.
$\Box$\\

In particular, we also note that in the above, $\Gamma_X^1-\Gamma_X^2$, when
inserted into another graph, produces a result with a bidegree reduced
by one unit compared to the insertion of either $\Gamma_X^1$ or
$\Gamma_X^2$ alone.

This is not yet the desired result, as in our case
we have to compare $\Gamma^1_{X_1}$ with $\Gamma^2_{X_2}$,
where $X_1$ is a collection of subgraphs in which all vertex subgraphs
are of type $\Gamma^1$, and  $X_2$ is the same collection of subgraphs
apart from  the replacement
$\Gamma^1\to\Gamma^2$ for all vertex subgraphs.

Any graph of type $\Gamma^1_{X_1}$ or $\Gamma^2_{X_2}$,
which itself can contain subgraphs $X_i$ of these varying types of vertex
corrections plus self-energy subgraphs,
can now be expressed in terms of the other.
Similarly, this holds for these vertex subgraphs of either type,
on the expense of
generating extra terms of reduced bidegree
$$
{\rm bid}(\Gamma^1_{X_1}-\Gamma^2_{X_2})<{\rm bid}(\Gamma^1_{X_1})
={\rm bid}(\Gamma^2_{X_2}),
$$
which involve differences $\Gamma_{X_i}^1-\Gamma_{X_i}^2$ for appropriate
$X_i$. Hence, under the above assumption of monotonic increase of the
transcendental weight with the bidegree, we get upon iterating such
insertions
\begin{prop}
The coefficient of the term of maximal transcendental weight is
the same for  ${\rm res}(\Gamma_{X_1}^1)$ and ${\rm res}(\Gamma_{X_2}^2)$.
\end{prop}
Here, $X_1,X_2$ are related, as above.

This explains immediately our results as a look at the functions
$F_{a,b}$, and hence the corresponding evaluations of our subdivergent
graphs, shows that they fulfill the required assumptions of monotonic
increase of transcendental weight, which was completely determined
from the appearances of the Riemann $\zeta$-function in our simple
examples. Note that the factorizations into two-point functions and
the absence of all other primitive graphs apart from one-loop
functions were the two main simplifications which enabled us to
satisfy the assumption.

The study to what extent a sensible transcendental weight can be
established in general will be a topic of future work. Any
sensible answer we will expect to deliver the same permutation
invariance of the residue as reported here.

\section*{Acknowledgements}

Isabella Bierenbaum acknowledges support by the {\em Graduiertenkolleg Eichtheorien - Experimentelle Tests und theoretische Grundlagen} at Mainz University.

\end{document}